%% file: paper.tex
\documentclass[10pt,sigconf,noacm=true,authorversion=true]{acmart}
\usepackage{pdflscape}

\usepackage{pifont}
\newcommand{\cmark}{\ding{51}}
\newcommand{\xmark}{\ding{55}}

\setcopyright{none}
\renewcommand\footnotetextcopyrightpermission[1]{} % removes footnote with conference information in first column
\usepackage[font=small,format=plain,labelfont=bf,textfont=it,skip=0pt]{caption}
\renewcommand{\paragraph}[1]{\vspace*{0.03in}\noindent\textbf{#1}}

\makeatletter
\def\subsubsection{\@startsection{subsubsection}{3}%
  \z@{.5\linespacing\@plus.7\linespacing}{.1\linespacing}%
    {\normalfont\itshape}}
    \makeatother

\title{Feature Extraction for Novelty Detection \\ in Network Traffic}
\author{Kun Yang}
\affiliation{Columbia University}
\author{Nick Feamster}
\affiliation{University of Chicago}
\author{Samory Kpotufe}
\affiliation{Columbia University}

\input{preamble}

\begin{document}
\begin{sloppypar}
\input{abstract}

\maketitle
\pagestyle{plain} % removes running headers

\input{intro}

\input{related}
\input{representation}
\input{metrics}

\input{system}

\input{setup}

\balance\input{results}

\balance\input{conclusion}

\microtypesetup{protrusion=false}

 \newpage
 \bibliographystyle{abbrv}
 \balance\bibliography{reference}

\vfill
\pagebreak
\input{appendix}

\end{sloppypar}
\end{document}

%% file: preamble.tex
\usepackage{url}            % simple URL typesetting
\usepackage{hyperref}
\usepackage{nicefrac}       % compact symbols for 1/2, etc.
\usepackage{microtype}      % microtypography
\usepackage{lipsum}
\usepackage{balance}
\usepackage{babel}
\usepackage{amstext}

\usepackage{booktabs}       % professional-quality tables
\usepackage{array}
\usepackage{multirow}
\usepackage{colortbl}
\usepackage{placeins}
\usepackage{float}
\usepackage{tabulary}
\usepackage{lscape}
\usepackage{rotating}
\usepackage{makecell}
\usepackage[usestackEOL]{stackengine}
\usepackage{longtable}
\usepackage{enumitem}
\usepackage[flushleft]{threeparttable}
\usepackage{graphicx}
\graphicspath{ {./figures/} }
\usepackage{subcaption}

\usepackage{algorithmic}
\usepackage[ruled, lined]{algorithm2e}
\usepackage{pifont}% http://ctan.org/pkg/pifont
\newcommand{\IAT}{$IAT$}
\newcommand{\SIZE}{$SIZE$}

\newcommand{\Cell}[1]{\begin{tabular}{@{}c}{\Centerstack[c]{#1}}\end{tabular}}

\definecolor{Gray}{gray}{0.9}
\newcolumntype{g}{>{\columncolor{Gray}}c}

\newlength{\offsetpage}
{\end{adjustwidth}}

\makeatletter
\DeclareRobustCommand*\cal{\@fontswitch\relax\mathcal}
\makeatother

\usepackage{listings}
\usepackage{xcolor}

\definecolor{codegreen}{rgb}{0,0.6,0}
\definecolor{codegray}{rgb}{0.5,0.5,0.5}
\definecolor{codepurple}{rgb}{0.58,0,0.82}
\definecolor{backcolour}{rgb}{0.95,0.95,0.92}

\lstdefinestyle{pstyle}{
    basicstyle=\ttfamily\scriptsize,
    backgroundcolor=\color{backcolour},   
    commentstyle=\color{codegreen},
    keywordstyle=\color{magenta},
    numberstyle=\tiny\color{codegray},
    stringstyle=\color{codepurple},
    basicstyle=\ttfamily\footnotesize,
    breakatwhitespace=false,         
    breaklines=true,                 
    captionpos=b,                    
    keepspaces=true,                 
    numbers=left,                    
    numbersep=5pt,                  
    showspaces=false,                
    showstringspaces=false,
    showtabs=false,                  
    tabsize=2
}

%% file: abstract.tex
\begin{abstract}

    Data representation plays a critical role in the performance of novelty
    detection (or ``anomaly detection'') methods in machine learning. The data
    representation of network traffic often determines the effectiveness of
    these models as much as the model itself.  The wide range of novel events
    that network operators need to detect (e.g., attacks, malware, new
    applications, changes in traffic demands) introduces the possibility for a
    broad range of possible models and data representations.  In each
    scenario, practitioners must spend significant effort extracting and
    engineering features that are most predictive for that situation or
    application. While anomaly detection is well-studied in computer
    networking,  much existing work develops specific models that presume a
    particular representation---often IPFIX/NetFlow.  Yet, other
    representations may result in higher model accuracy, and the rise of
    programmable networks now makes it more practical to explore a broader
    range of representations.  To facilitate such exploration, we develop a
    systematic framework, open-source toolkit, and public Python library that
    makes it both possible and easy to extract and generate features from
    network traffic and perform and end-to-end evaluation of these
    representations across most prevalent modern novelty detection models.  We
    first develop and publicly release an open-source tool, an accompanying
    Python library (NetML), and end-to-end pipeline for novelty detection in
    network traffic. Second, we apply this tool to five different novelty
    detection problems in networking, across a range of scenarios from attack
    detection to novel device detection. Our findings general insights and
    guidelines concerning which features appear to be more appropriate for
    particular situations.   
\end{abstract}

%% file: intro.tex
%%%%%%%%%%%%%%%%%%%%%%%%%%%%%%%%%%%%%%%%%%%%%%%%%%%%%%%%%%%%%%%%%%%%
% Introduction 
%%%%%%%%%%%%%%%%%%%%%%%%%%%%%%%%%%%%%%%%%%%%%%%%%%%%%%%%%%%%%%%%%%%%

\section{Introduction}

Novelty detection (often also referred to as anomaly detection) is a common
network management task: Network operators commonly need to detect and respond
to a wide variety of new or unusual events, such as malicious attacks,
failures, misconfigurations, changes in traffic demands, the introduction of
new devices to the network, and so forth. 
These events can be
indicative of many types of noteworthy events, including network or physical
security incidents, device or infrastructure malfunctions, even behavioral
changes in the users of the devices that might be indicative of abnormal human
conditions or behaviors. 

The goal of detecting such new, unseen events falls under the general category
of {\em novelty detection} (also known as {\em anomaly detection}) problems in
machine learning, for which there are many possible procedures, including
Gaussian mixture models (GMM), one-class support vector machines (OCSVM),
kernel density estimation (KDE), and others. The effectiveness of any of these
procedures---and, in fact, for {\em any} machine learning approach---depends
fundamentally on how the data are represented as input features to the model.
Often the choice of features and how they are represented---the process of
{\em feature extraction}---determines the accuracy and effectiveness of the
approach than the choice of model. 

Representations of network traffic that are most effective for novelty
detection are typically {\em a priori} unclear.  Representation choices
typically involve various statistics concerning traffic volumes, packet
sampling rates, and aggregation intervals. Yet, which of these transformations
retain predictive information is typically initially unknown.  While novelty
detection is a problem for many problems in networking, and for network
traffic in general, the problem is particularly challenging for Internet of
Things (IoT) traffic, given the wide range of device types, vendors, traffic
patterns, interaction modes, and failure scenarios across devices.  One
example challenge is choosing the granularity on which to aggregate a feature:
For instance, if we decide to measure and represent the number of packets per
second as a feature, then changes in traffic patterns at the millisecond level
might go undetected; on the other hand, recording at this granularity creates
higher dimensional observations introducing high-dimensional data that creates
challenges with both scale and accuracy.  (The so-called \emph{curse of
dimensionality} in machine learning is the observed fact that the higher the
dimension of the input, the harder it is to achieve good accuracy; in
particular, we might then require an amount of training data exponential as a
function of dimension.)

This paper explores how the process of feature extraction from raw network
traffic affects the accuracy of a variety of novelty detection methods for a
range of novelty detection problems in network traffic. To facilitate the
reproducibility of these results and further research on novelty detection
from network traffic, we have released our feature extraction methods as an
open-source library and a publicly available Python package on PyPi.
We aim to explore how different data representations for
network traffic affect novelty detection rates, across (1)~different types of
novelty detection problems; (2)~different types and classes of network
devices; (3)~a range of unsupervised machine learning models.  We focus on
identifying novel network flows, a common, intuitive unit of identification
that, at a high-level, represents a single exchange of data between two
network endpoints. The ability to identify a novel flow lends itself to
detecting other types of novelty; for example, a device that generates novel
traffic flows may itself be considered novel. 

There are many possible ways to represent raw traffic traces (i.e., packets)
as inputs. The data itself can be aggregated on various time windows, elided,
transformed into different bases, sampled, and so forth.  We focus on
representations of network traffic data that do not incorporate semantics from
the header itself, such as IP addresses, port numbers, sequence numbers, or
other specific values in the packet headers themselves. Representations that
are agnostic to specific values in the headers ensures that the resulting
models are trained to recognize novel behavior regardless of the {\em
specific} source, destination, application, or location in the network. In
contrast to past work in this area, which has typically focused on a single
problem in novelty detection using a single dataset and a chosen procedure,
our focus is on studying the applicability of representation, we aim for {\em
generality} and reproducibility of results across a range of problems,
studying the applicability of different representations across a range of
problems.

When evaluating different feature representations across models and problem
domains, we design methods and metrics that result in comparisons that are as
fair as possible. We report the area under the ROC curve (AUC) for each
approach, scenario, and dataset.  Because machine learning procedures are very
sensitive to hyperparameter choices (e.g., bandwidth choice in OCSVMs, or the
number of dimensions in PCA), we consider both the best possible choices of
such hyperparameters (which unfortunately are hard to find in practice without
labeled data), and default choices that might be made in practice based on
rules-of-thumb. To ensure an apples-to-apples comparison across procedures, we
ensure that all data representations result in feature vectors of a similar
order in dimension, in the range of 10 to 30 dimensions.  Our goal is {\em
not} to identify the best outlier detection model but rather to demonstrate
that certain feature representations are effective across models. The public,
open-source library that we have released makes it possible to transform raw
packet captures into the representations we explore in this paper and explore
the accuracy of outlier detection across many families of models.  

Our exploration of different novelty detection procedures across a range of
networking problems yields the following findings:
\begin{itemize}[leftmargin=*]
\itemsep=-1pt
\item {\bf No single data representation works best across all devices and
    models.} The lack of a singe best representation underscores the
        importance of exploring a wide range of possible features and
        representations, which we have made possible through the release of an
        open-source toolkit and Python library.
    \item {\bf Incorporating metadata such as packet size and header information (e.g.,
        IP TTL, TCP flags) significantly improves the accuracy
        of novelty detection, across a variety of models and outlier detection
        techniques.} This result has significant implications on past work in
        the Internet measurement and security areas, which have focused
        predominantly on representations such as IPFIX (e.g., NetFlow), which
        do not incorporate these features.
    \item {\bf Frequency domain representations often have little effect on
        model accuracy.} Contrary to expectations, performing a Fourier
        transform on the raw network packet traces generally has minimal
        effect on model accuracy, regardless of the representation of the data
        in the time domain.
\end{itemize}
These results hold generally across all families of models we explore, for a
wide range of anomaly detection models, from physical device anomalies to
security events. These results represent a promising step forward for anomaly
detection, as they demonstrate that packet header values---which are simple to
extract and visible even when traffic is encrypted---can be sufficient for
anomaly detection, independent of model, subject to the appropriate
transformations on their representation.
To make these transformations possible, to facilitate the reproducibility of
our results as well as follow-on work that can make use of the packet
transformation capabilities we have developed and evaluated in this work, we
have published our source code as an open-source software
library~\cite{itod-code}.

The rest of the paper is organized as follows. Section~\ref{sec:related}
discusses related work, and Section~\ref{sec:data_representations} outlines
the common ways of representing network traffic flows in the literature. It is
difficult to be completely exhaustive, yet we focus on the most established
approaches for representing traffic data both in general network anomaly
detection and in the growing literature on novelty detection for IoT,
comparing tuning and configuration and tuning decisions that might be made in
practice.  Section~\ref{sec:ml_procedures} outlines the machine learning
models and performance metrics that we use for our comparative analysis, 
the metrics that we use to evaluate these models.
and the setup of our experiments,
how we selected hyperparameters for various novelty detection
models, and the datasets that we used in our evaluation and how we
processed those datasets. Section~\ref{sec:results} describes the results from
our experiments, and Section~\ref{sec:conclusion} concludes with a summary and
directions for future work.

%% file: related.tex
\section{Related Work}\label{sec:related}

Novelty detection in computer networking is a well-explored topic. Past
research has explored relatively standard techniques, such as Principal
Component Analysis~(PCA), for anomaly detection in network traffic. Clustering
approaches, including PCA-based clustering approaches (e.g., spectral
clustering) has also been applied to network traffic to detect anomalous
behavior. In contrast, relatively less attention has been paid to a broader
range of more modern anomaly detection models that we explore in this
paper---many of which outperform PCA. Previous work has also applied various
techniques towards novelty detection using a broad range of data
representations, without comparing the (potentially significant) effect that
data representation can have on model accuracy. We briefly survey related work
in both unsupervised learning methods for novelty detection in networking, as
well as various data representations that past research has used for novelty
detection.

\paragraph{Principal Component Analysis.}
Previous work has focused on the use of unsupervised methods for novelty
detection---sometimes also referred to as ``anomaly detection''. A significant
amount of work on anomaly detection in network traffic has focused on
Principal Component Analysis~(PCA).  PCA has been used detect anomalous
traffic flows in network
backbones~\cite{lakhina2005mining,lakhina2004diagnosing,lakhina2004structural}.
Subsequent work observed that the effectiveness of PCA for network anomaly
detection has a significant dependence on how network traffic data is
represented~\cite{ringberg2007sensitivity}. This work found that PCA is
sensitive to various aspects of data representation, including the level
of aggregation that is performed in the original traffic traces. This work and
and others also demonstrated that PCA-based anomaly detection is not robust to
noise~\cite{rubinstein2009stealthy}. More recently, PCA has been applied to
anomaly detection for in IoT network traffic~\cite{hoang2018pca}.

Despite the significant attention devoted to PCA for anomaly detection,
relatively little attention has been paid to the much broader family of
anomaly detection that we study in this paper, including models that can learn
non-linear decision boundaries for detecting novel activity (e.g., one-class
SVM, auto-encoder) as we explore in this paper.
Furthermore, as previously discussed, the {\em representation} of the traffic
data is often at least as important as the choice of model; Ringberg {\em et
al.} first noted the sensitivities of PCA to data representation in
2007~\cite{ringberg2007sensitivity}. Despite the importance of data
representation on detection accuracy, there has been little study of this
problem. To our knowledge, our work is the first to explore a wide range of
possibilities for representing network traffic, for a large collection of
datasets and models.

\paragraph{Clustering Techniques.}
Clustering techniques have been used to detect certain types of malicious
network activity such as botnets~\cite{gu2008botminer} and web-based
malware~\cite{perdisci2010behavioral}. These works apply hierarchical
clustering techniques to identify groups of hosts that are behaving in a
similar fashion, thereby indicating potential botnet membership.  Other
clustering techniques, such as spectral clustering, have been applied to email
traffic to detect spammers~\cite{ramachandran2007filtering}.  These methods
are generally specific to the detection of botnet behavior, which tends to
cluster well; additionally, this past research uses a variety of traces (e.g.,
IPFIX/NetFlow, HTTP traces, email logs).  Our work focuses instead on novelty
detection in raw packet traces, across many models and novelty detection
problems.

\paragraph{Metric Learning and Representation Learning.}
The importance of data representation for model accuracy has fostered many
subfields such as \emph{metric learning} and \emph{representation learning},
which aim to \emph{automatically} find the right representation to be coupled
with given procedures, so far with mixed results. These techniques have
achieved much success in application areas such as computer vision and speech
processing, yet no such automated approaches exist in the fledgling
application area of novelty detection for network traffic, particularly for
networked IoT devices. A main aim of this work is to bring attention to the
importance of data representation, and identify representations that work well
in general while meeting practical requirements for succinct, efficient representations.

\paragraph{Data Representations for Anomaly Detection.}
Previous work has explored the use of various subsets of network traffic
features for detecting anomalies and security incidents in network traffic.
Lakshari {\em et al.} used flow direction, interarrival time, and
packet sizes as features~\cite{lashkari2018toward}; Mukar {\em et al.}
used packet inter-arrival time~\cite{kumar2018novel}, and Mirsky {\em et al.}
used flow statistics over fixed time intervals, or windows, as
feautres~\cite{mirsky2018kitsune}. Other researchers have explored
various combinations of these features for IoT anomaly
detection~\cite{aljawarneh2018garuda, doshi2018machine,
meidan2018n,bhatia2019unsupervised,thamilarasu2019towards},
specifically, as well as for detection of distributed denial of service (DDoS)
attacks~\cite{fouladi2016frequency,lima2019smart}.
Table~\ref{tab:representations} summarizes this past work, and various data
representations that each has used in developing novelty detection
models. All of this past research differs significantly from our work, in that
each work typically applies {\em one} representation and model to a specific
problem. In contrast, our work attempts to explore all of these
representations across a wide range of models, evaluating how well specific
representations work for different types of novelty detection problems. To our
knowledge, our work is the first to take such a holistic approach to novelty
detection in network traffic, and the first to explore a holistic approach to
a wide range of settings and scenarios in IoT.

\if 0
\midrule
    Lakshari {\em et al.}~\cite{lashkari2018toward} & \cmark & \cmark& \cmark & \xmark & \xmark & \xmark  \\
\midrule
    Mukar {\em et al.}~\cite{kumar2018novel} & \xmark & \cmark& \xmark & \xmark & \xmark & \xmark  \\
    Mirsky {\em et al.}~\cite{mirsky2018kitsune} &\xmark & \xmark& \xmark & \xmark & \cmark & \cmark  \\
\midrule
    \multicolumn{2}{l}{\textbf{IoT Anomaly Detection}} & \multicolumn{4}{l}{} &\\ 
\midrule
    Al Jawarneh {\em et al.}\cite{aljawarneh2018garuda} & \cmark & \xmark& \cmark & \xmark & \xmark & \xmark  \\
\midrule
    Doshi {\em et al.}~\cite{doshi2018machine} & \xmark & \cmark& \cmark & \xmark & \xmark & \xmark  \\
\midrule
    Meidan {\em et al.}~\cite{meidan2018n} &\xmark & \cmark& \xmark & \xmark & \cmark & \cmark  \\
\midrule
    Bhatia {\em et al.}~\cite{bhatia2019unsupervised} & \xmark & \xmark& \cmark & \xmark & \cmark & \xmark  \\
\midrule
    Thamilarasu {\em et al.}~\cite{thamilarasu2019towards} & \xmark & \xmark& \cmark & \xmark & \xmark & \xmark  \\
\midrule
    \multicolumn{2}{l}{\textbf{DDoS Detection}} & \multicolumn{4}{l}{} &\\ 
\midrule
    Fouladi {\em et al.}~\cite{fouladi2016frequency} &\xmark & \xmark& \xmark & \cmark & \xmark & \xmark  \\
\midrule
    Lima {\em et al.}~\cite{lima2019smart} &\xmark & \xmark& \cmark & \xmark & \xmark & \xmark  \\

\fi

%% file: representation.tex
%%%%%%%%%%%%%%%%%%%%%%%%%%%%%%%%%%%%%%%%%%%%%%%%%%%%%%%%%%%%%%%%%%%%
% IAT and ML 
%%%%%%%%%%%%%%%%%%%%%%%%%%%%%%%%%%%%%%%%%%%%%%%%%%%%%%%%%%%%%%%%%%%%

\section{Representing Network Traffic}
\label{sec:data_representations}
\label{sec:data}

Machine learning procedures require for the most part, \emph{feature vectors}
of the same dimension, i.e., every flow has to be represented as a vector of a
fixed number of carefully chosen features, specifically information extracted
from the flow.  

\paragraph{Flows.} 
We consider various feature representations of \emph{forward
flows}, where a forward flow aggregates
all packets, with a same \emph{five-tuple} identifier, sent from a device
being monitored to another device or server. The five-tuple identifier
consists of source IP address, source port, destination IP address,
destination port, and protocol. We consider the following representations of
flows:

\begin{table}[t!]
\hspace*{-0.25in}
\centering
\begin{scriptsize}
\begin{tabular}{|lcccccc|}
\toprule
Reference & Duration & IAT & SIZE & FFT & SAMP-NUM & SAMP-SIZE \\
\midrule
    \multicolumn{2}{l}{\textbf{Network Intrusion Detection}} & \multicolumn{4}{l}{} & \\
\midrule
    Lakshari {\em et al.}~\cite{lashkari2018toward} & \cmark & \cmark& \cmark & \xmark & \xmark & \xmark  \\
\midrule
    Mukar {\em et al.}~\cite{kumar2018novel} & \xmark & \cmark& \xmark & \xmark & \xmark & \xmark  \\
\midrule
    Mirsky {\em et al.}~\cite{mirsky2018kitsune} &\xmark & \xmark& \xmark & \xmark & \cmark & \cmark  \\
\midrule
    \multicolumn{2}{l}{\textbf{IoT Anomaly Detection}} & \multicolumn{4}{l}{} &\\ 
\midrule
    Al Jawarneh {\em et al.}\cite{aljawarneh2018garuda} & \cmark & \xmark& \cmark & \xmark & \xmark & \xmark  \\
\midrule
    Doshi {\em et al.}~\cite{doshi2018machine} & \xmark & \cmark& \cmark & \xmark & \xmark & \xmark  \\
\midrule
    Meidan {\em et al.}~\cite{meidan2018n} &\xmark & \cmark& \xmark & \xmark & \cmark & \cmark  \\
\midrule
    Bhatia {\em et al.}~\cite{bhatia2019unsupervised} & \xmark & \xmark& \cmark & \xmark & \cmark & \xmark  \\
\midrule
    Thamilarasu {\em et al.}~\cite{thamilarasu2019towards} & \xmark & \xmark& \cmark & \xmark & \xmark & \xmark  \\
\midrule
    \multicolumn{2}{l}{\textbf{DDoS Detection}} & \multicolumn{4}{l}{} &\\ 
\midrule
    Fouladi {\em et al.}~\cite{fouladi2016frequency} &\xmark & \xmark& \xmark & \cmark & \xmark & \xmark  \\
\midrule
    Lima {\em et al.}~\cite{lima2019smart} &\xmark & \xmark& \cmark & \xmark & \xmark & \xmark  \\
\bottomrule
\end{tabular}
\end{scriptsize}
\caption{Commonly used data representations for novelty detection.}\label{tab:representations}
\label{features}
\end{table}

\begin{itemize}[leftmargin=0.2in]
    \item {\bf STATS:} A flow is represented as a set of statistical
        quantities.
        We choose ten of the most common such statistics in the literature (see
        i.e, \cite{Moore2005}): flow duration, number of packets sent per
        second, number of bytes per second, and various statistics on
        {packet sizes} within each flow: mean, standard deviation,
        inter-quartile range, minimum, and maximum.

\item {\bf SIZES:} A flow is represented as a timeseries of packet sizes in
    bytes, with one sample per packet. 

\item {\bf IAT:} A flow is represented as a timeseries of inter-arrival times
    between packets, i.e., elapsed time in seconds between any two packets in
        the flow. 

\item {\bf SAMP-NUM:} A flow is partitioned into small intervals of equal
    length $\delta t$, and the number of packets in each interval is recorded;
        thus a flow is represented as a timeseries of packet counts in small
        time intervals, with one sample per time interval. Here, $\delta t$ might be viewed as a choice of
        \emph{sampling rate} for the timeseries, hence the nomenclature. 

\item {\bf SAMP-SIZE:} A flow is partitioned into time intervals of equal
    length $\delta t$, and the total packet size (i.e., byte count) in each
        interval is recorded; thus, a flow is represented as a timeseries of
        byte counts in small time intervals, with one sample per time interval. 
\end{itemize} 
\noindent
Any of these timeseries representations might be
given in time or Fourier domain, in
which case we append \emph{FFT} to the representation name, i.e., IAT-FFT,
SAMP-NUM-FFT, and SAMP-SIZE-FFT.  

In all cases, we avoid any representations that rely on or encode specific
{\em values} of packet headers, such as the value of a source or destination
IP address or port number, the value of a protocol (e.g., TCP, UDP), and so
forth, to avoid developing models that would define normal behavior as
specific to a particular network attachment point (e.g., IP address or port).
We avoid training models based on these values for several reasons. Most
importantly, training a model based on IP address values would define anomaly
simply as any deviation in network destinations from those in the trained
model. A model trained on traffic traces collected in one part of the network
would not be able to perform general novelty detection when applied or
deployed on other parts of the network. The model would also fail to detect
novel behaviors from the same IP addresses over time---as would be the case if
a device were compromised or simply changed its behavior as result of changes
in the environment or how users interacted with the device over time.  For
similar reasons, we avoid relying on port numbers. On the one hand, a
significant amount of traffic---both normal and novel---traverses common ports
(e.g., port 443, or HTTPS); on the other hand, {\em client} traffic
originates from ports that change (e.g., increment) all of the time, and thus
defining normal traffic based on client ports would result in a significant
amount of false positives simply based on the normal behavior of clients
originating connections from different ports over time.

When applying different machine learning models for novelty detection, we must
ensure that all of the above representations result in feature vectors of the
same dimension for any given device being monitored. This is automatically the
case for STATS, but not for the other representations, in light of variations
in flow durations and lengths (number of packets) for a given device.  To
ensure that each sample has the same dimension, we select a fixed time
duration for all flows corresponding to the 90th percentile of all flow
lengths or durations for a device. We use this duration so as to capture the
typical behavior of most device instances of a particular type. Any definition
of typicality may be appropriate here, depending on the desired compactness
and complexity of the corresponding representation. Using the 90th percentile
of all flow durations results in a comparable number of dimensions (typically
10--30) for most of the devices that we consider in this study.  

%% file: metrics.tex
\section{Models and Metrics}
\label{sec:metrics}
\label{sec:ml_procedures}
% After describing and explaining data representations in Section  \ref{sec:data_representations}, we start to

We describe the novelty detection procedures that we evaluate in our study, as
well as the performance metrics that we use to compare each of these
procedures.
We then
describe the datasets we use to evaluate each novelty detection method,
including how we process each of these datasets and select parameters for each
corresponding feature representation.
Finally, we describe the process by which we choose hyperparameters for each of
the novelty detection procedures that we evaluate in this study. 

\subsection{Novelty Detection Models} 

We consider popular novelty detection procedures from machine learning and
statistics, approaches based on six models: 
\begin{itemize}
    \item {One-Class Support Vector Machines} (OCSVM), 
    \item {Isolation Forests} (IF),
    \item {Auto Encoders} (AE)
    \item {Kernel Density Estimation} (KDE)
    \item {Gaussian Mixture Models} (GMM), and
    \item {Principal Component Analysis} (PCA).
\end{itemize}
Important common aspects of all such detection procedures are as follows:

\begin{enumerate}[leftmargin=*]
\item \textbf{Data representation as features.} Novelty detection procedures invariably
    require data---network flows---represented as a vector $X \in \mathbb{R}^d$ of $d$ predictive features. 
    While many procedures operate directly on that feature representation (e.g.,
        IF, KDE, GMM), others (e.g., OCSVM, PCA, AE) remap vectors $X$ into
        refined representations where certain patterns---clusters of similar
        network activities---might be more evident. 

\item \textbf{Training and testing.} Models are trained in a first phase
    using $n$ training data points $\{X_i\}_{i=1}^n$ -- i.e., $n$ copies $X_i$
        of typical values of \emph{normal} features $X$ (e.g., each $X_i$
        representing a normal traffic flow), from which they produce a
        \emph{scoring function $S(X)$};  high scores $S(X)$ are expected to be
        typical of normal observations such as the training data points $X_i$,
        while low-scores would be indicative of the query $X$ being an outlier
        (novelty). Subsequently, in the so-called testing phase (or deployment
        phase), a threshold $t$ might be chosen so that any query $X$ is
        flagged as \emph{novelty} whenever $S(X) < t$. 
\end{enumerate} 
\noindent
In practical settings, the right choice of threshold can be difficult since
the available training data is assumed to be all \emph{normal}, making it
difficult 
to assess \emph{false negative rates}. Instead, one might choose a threshold
based on the \emph{false positive rates} the application can tolerate, which
can be estimated for each choice of $t$ based on the \emph{normal data} (i.e.,
data known to not have positive label). Related issues are discussed in
Section \ref{sec:performancemetric} on performance metrics.

\paragraph{Normality scores in ML:} In KDE or GMM, the training data is used to estimate the \emph{density} function $f$ of the data in $\mathbb{R}^d$ (that is their spatial distribution), where by definition $f(x)$ is low in those regions of space that have little data. This density would then serve as a scoring function $S(X) \doteq f(X)$. IF works similarly by first partitioning the space $\mathbb{R}^d$ into ensembles of high and low density regions, and flagging any query $X$ as a novelty if it lies in a low density region. 

On the other hand, PCA and AE work by identifying a lower-dimensional space ${\cal X}$ embedded in the representation space $\mathbb{R}^d$ on which most of the training datapoints $X_i$ appear to lie. In PCA $\cal X$ is a linear space, while in AE it can be highly nonlinear (using nonlinear activation functions in a neural network). The score $S(X)$ then denotes how \emph{close} $X$ is to $\cal X$ with lowest scores for points farthest from $\cal X$. %Note that these might also be re-interpreted as low-density approaches, since $\mathbb{R}^d \setminus \cal X$ might be viewed as a low density region of $\mathbb{R}^d$. 

OCSVM stands farther apart as it proceeds by first remapping all $X_i$'s -- through a transformation $\Phi(X_i)$ into a higher dimensional feature space $\cal H$ (possibly of infinite dimension) and then essentially identifying regions of high density in $\cal H$, typically by fitting a hyperplane separating the embedded normal data from the origin $0\in \cal H$. 
The underlying assumption is that patterns might be more visible in $\cal H$, for instance, all $\Phi (X_i)$ might cluster together into a \emph{normal} region (typically a half space in $\cal H$, or sometimes an enclosing ball; the score $S(X)$ is then lower for queries $X$ that map as $\Phi(X)$ outside the \emph{normal} region of $\cal H$. 

Although we analyze all the above described procedures, we restrict attention to a few representative such ML procedures in the main body of the paper (see Section \ref{sec:results}), while all additional substantiating results can be found in the appendix. 

\subsection{Performance Metrics}\label{sec:performancemetric}
% AUC (obtained by four machine learning approaches: GMM, OCSVM, KDE, and AE)
\paragraph{Area Under the Curve (AUC):}
As explained above, detection procedures flag a new query $X$ as novel if it
scores below a certain threshold $t$ (i.e., if $S(X) < t$). Now notice a
tension: the higher the threshold $t$, the more likely it is that all novel
$X$ are correctly detected, but unfortunately, the more likely it is also that
normal datapoints are flagged as novel (a false alarm). Thus, the best
performing approaches---combination of data representation and choice of
detection procedure---are those that alleviate this tension, performing
accurate detection while minimizing false positives.

Such tradeoffs are well captured by a Receiver Operating Characteristic (ROC)
curve, which plots the detection rate against the false alarm rate as $t$ is varied from small to large; a large area under the curve (AUC) indicates that good tradeoffs are possible under the given detection approach. We therefore adopt AUC as a sensible measure of performance. 

\paragraph{Sensitivity to Hyperparameters:} All detection procedures come with hyperparameters, i.e., configuration choices that can greatly affect performance. Important hyperparameters are e.g., bandwidth in KDE or OSCVM, number of partitions in IF, number of components in GMM, embedding dimension in PCA and AE. We therefore have to carefully pick such parameters in practice, although the right approach remains unclear, especially in \emph{unsupervised learning} problems such as novelty detection where we typically do not have a labeled validation set to evaluate choices. 

{Since hyperparameter tuning in unsupervised learning remains the subject of ongoing research in Machine Learning and Statistics, \emph{we decided to present results for the best choices of parameters}, picked as those choices that result in the best performance on a separate validation set, and used to report results on an independent test set. The tuning parameters for each method are described in Section \ref{sec:hyperparamTuning} below, while Appendix \ref{tab:train_test_sets} gives data sizes (with validation kept around 1/4 test sizes). }

More configuration details such as range or parameters are given in Section \ref{sec:hyperparamTuning}. Other configuration choices based on common rule-of-thumbs are covered in more detail in the appendix along with relevant results.

%% file: system.tex
\section{NetML: A Novelty Detection Library for Network Traffic}

We have designed, implemented, and released a public, open-source Python
library, {\tt netml}, that takes network packet traces as input, transforms
these packet traces into the various data representations as outlined in
Section~\ref{sec:metrics}. We have released the open-source code for the
library and packaged the library on PyPi for easy installation.

\begin{figure}[t]
\begin{lstlisting}[language=Python, caption=From packet capture to features.,label=data]
from netml.pparser.parser import PCAP
from netml.utils.tool import dump_data

pcap = PCAP(
    'data.pcap',
    flow_ptks_thres=2,
    random_state=42,
    verbose=10,
)

# extract flows from pcap
pcap.pcap2flows(q_interval=0.9)

# label each flow (optional)
pcap.label_flows(label_file='labels.csv')

# extract features from each flow via IAT
pcap.flow2features('IAT', fft=False, header=False)

# dump data to disk
dump_data((pcap.features, pcap.labels),
            out_file='IAT-features.dat')
\end{lstlisting}
\end{figure}

The {\tt netml} library is written in Python and contains two sub-modules:
(1)~a {\tt pcap} parser to produce flows using Scapy/dpkt; (2)~a novelty
detection module that applies the novelty detection algorithms we outlined in
Section~\ref{sec:metrics} to the resulting flow-based features. The {\tt
netml} functionality can be incorporated as a Python library or invoked
directly from the command line.
The entire library consists of approximately 11,000 lines of Python, with the
packet parsing module comprising approximately 3,500 lines of Python, the
implementations of various novelty detection modules comprising another 3,000
lines, another 3,000 lines in support of the {\tt netml} command-line utility,
and the rest of the code in support of various utilities (e.g., testing).

Listing~\ref{data} shows the basic functions to extract features from
traffic flows. Line~12 extracts flow from the packet capture and Line~15 labels
time intervals in the flow according to whether those are normal or anomalous.
The {\tt flows2features} function at line~18 extracts features from the flows
in the pcap according to the specified parameters. The library currently
supports all of the representations outlined in
Section~\ref{sec:data}.

\begin{figure}[t]
\begin{scriptsize}
\begin{lstlisting}[language=Python, caption=Training and testing models.,label=model]
from netml.ndm.model import MODEL
from netml.ndm.ocsvm import OCSVM
from netml.utils.tool import dump_data, load_data

# load data
(features, labels) = load_data('IAT-features.dat')

# split train and test sets
(
    features_train,
    features_test,
    labels_train,
    labels_test,
) = train_test_split(features, labels, 
                        test_size=0.2)

# create detection model
ocsvm = OCSVM(kernel='rbf', nu=0.5)
ocsvm.name = 'OCSVM'
ndm = MODEL(ocsvm, score_metric='auc')

# train the model from the train set
ndm.train(features_train)

# evaluate the trained model
ndm.test(features_test, labels_test)
\end{lstlisting}\label{lst:model}
\end{scriptsize}
\end{figure}

Listing~\ref{model} shows an example of how the library can be used to
train the model. As shown in the figure, model training and applications of
the model can be performed using function calls that are analogous to those in
the familiar scikit-learn API. Line~18 instantiates a particular novelty
detection model (in this case, OCSVM with a radial basis function kernel),
and Line~20 creates the model itself which can then be invoked with familiar
{\tt train} and {\tt test} functions. While it is not shown in the figure, the
module also records both accuracy and performance statistics that can then be
written to a data structure or to persistent storage.

%% file: setup.tex
\section{Experiment Setup}

We now describe the data that we used to evaluate the collection of novelty
detection algorithms and representations described in the previous section.

%%%%%%%%%%%%%%%%%%%%%%%%%%%%%%%%%%%%%%%%%%%%%%%%%%%%%%%%%%%%%%%%%%%%
% Experiment
%%%%%%%%%%%%%%%%%%%%%%%%%%%%%%%%%%%%%%%%%%%%%%%%%%%%%%%%%%%%%%%%%%%%
\subsection{Data Collection and Preprocessing}

We consider a combination of publicly available traffic traces and
traces collected on private consumer IoT devices. We aim to evaluate
representative set of devices, from multi-purpose devices such as laptop PCs,
and Google Home, to less complex electronics and appliances such as smart
cameras and TVs. Furthermore we aim at a representative set of \emph{novelties}, from
benign novelties (new activity, or a new device type), to novelties due to
malicious activities (DDoS attack).  Table \ref{dataset_sources} describes
these datasets, and types of novelty being detected. 

\input{results-table}

\begin{table*}[t!]
\centering
\begin{scriptsize}
% \begin{tabular}{|p{1.5cm}|p{6.5cm}|p{1.2cm}|p{2.5cm}|} 
\begin{tabular}{|p{2cm}|p{4.0cm}|p{8.5cm}|p{1.5cm}|} 
\toprule
Reference &Device  & Description & Novelty  \\
\hline
\Cell{UNB IDS~\cite{cicids2017}\\(5 datasets)}  & 
Five PCs (noted as PC1, PC2, PC3, PC4, and PC5, respectively) &Normal and DDoS attack traces from personal computers (PC), which are chosen from Monday trace and Friday trace.   &  DDoS attack\\ 
\hline
\Cell{CTU IoT~\cite{ctu2019iot}\\(1 dataset)}  & Two Raspberry Pis &
 Bitcoin-Mining and Botnet traffic traces generated by two Raspberry Pis; we aim to distinguish the two devices by their attack traces (i.e., Mirai and CoinMiner); hence we use one generated CoinMiner as \emph{normal}, the other generated Mirai as \emph{novel}.   &  Novel (infected) device  \\ 
\hline
\Cell{MAWI~\cite{mawi2019normal}\\(1 dataset)} & Two PCs & Normal traffic generated by two PCs, which are collected on Dec. 07, 2019; we use one PC (whose IP address is 202.171.168.50) as \emph{normal}, the other (whose IP address is 202.4.27.109) as \emph{novel}.    &   Novel (normal) device \\ 
\hline
\multirow{5}{*}{\Cell{Lab IoT\\(5 datasets)}} & A smart TV and a wireless router (noted as TV\&RT)  &  Data is generated by a smart TV and a wireless router in a private lab environment.  
We collect two kinds of normal traffic generated by a smart TV (whose IP address is 10.42.0.119) and a router (whose IP address is 10.42.0.1). the router traffic is labeled as  \emph{normal}, and the smart TV traffic is labeled as \emph{novel}. &
  Novel (normal) device  \\ 
\cline{2-4} & Google home (GHom), Samsung camera (SCam), Samsung fridge (SFrig), and Bose soundtouch (BSTch) 
 &Data traces are generated by IoT devices in a private lab environment.  Four smart-home devices, each with two types of traffic traces labeled as \emph{normal} when there is no human interaction, and \emph{novel} when being operated by a human. 
 
%  collected from four IoT devices. Each device has idle traffic collected during the idle of the device and active traffic collected during the period of some normal operations executed by a person. 
  & Novel activity  \\
\bottomrule
\end{tabular}
\end{scriptsize}
\caption{Datasets.}
\label{dataset_sources}
\end{table*}

\paragraph{Obtaining flows:}
We parse upstream flows from datasets in Table \ref{dataset_sources} using
    Scapy \cite{scapy}. Given that certain devices can have arbitrarily long
    flows, we truncate each flow from a given dataset to have duration at most
    that of the 90th upper-percentile of flow durations in the dataset.
    Henceforth, a \emph{flow} refers to these choices of flows involving
    truncation. 

\paragraph{Selecting feature representation dimension:}
Each flow is then represented using any of the feature-representation choices described in Section \ref{sec:data_representations}. Here, we have to ensure that, given a representation choice, e.g., IAT, 
all flows result in vectors of the same dimension $d$ for each dataset. We apply a similar approach to enforce such fixed dimension $d$, as outlined below. For a fixed dataset, let $d_0$ denote the 90th upper-percentile of number of packets per flow. 
\begin{itemize}[leftmargin=0.2in]
    \item STATS: automatically results in vectors of the same dimension $d$,
        equal to the number of statistics (on packet size) computed on a flow. 
    \item IAT, or SIZE: dimension is fixed to $d = d_0-1$ for IAT, and $d = d_0$ for size; namely, for long flows, we truncate the number of packets in a flow to the first $d_0$; for short flows where the number of packets is less than $d_0$, we append $0$'s to the IAT representation to arrive at $d$ features. 
    \item IAT + SIZE: just concatenates IAT and SIZE representations to dimension $d = 2d_0 -1$.  
    \item SAMP-NUM or SAMP-SIZE: dimension is fixed to $d = d_0-1$ as in the case of IAT. Now we pick 
    a fixed window size $\delta t$, and each flow is divided into up to $d$ windows of length $\delta t$; for flows of short duration, where $\delta t$ is large with respect to duration, we simply append $0$'s. 
    
    \emph{Choice of $\delta t$:} we start with candidate choices $\Delta
        \doteq \{t_f/d\}$, over durations $t_f$ of flows $f$ in a given
        dataset. We then consider 10 choices of $\delta_t$ in $\Delta$, each
        choice an $i$-th quantile of values in $\Delta$, for each $i \in \{10,
        20, 30, 40, 50, 60, 70, 80, 90, 95\}$. Thus, each such choice of $\delta t$ yields one (SAMP-NUM or SAMP-SIZE) representation for all flows; since we do not know which such representation (i.e. choice of $\delta t$) is best a priori\footnote{In fact we do not know how such choice might be automated in practice, as automating \emph{sampling rates} remains an largely open research problem.} we report the best AUC resulting out of all 10 choices of $\delta t$. 
\end{itemize}

Each of the above representations, besides STATS, can then be viewed as fixed length time series, and therefore admit a Fourier domain representation; the corresponding Fourier representation is then set to the same dimension $d$, i.e., we retain as many Fourier components as the original time series. 

The resulting representation dimensions for each dataset are reported in Table \ref{tab:feature_dimensions} in the appendix.

\subsection{Hyperparameter Settings}
\label{sec:hyperparamTuning}
All detection procedures are implemented in the scikit-learn Python package.
We now describe configuration choices, both for the best hyperparameter choice
(Optimal)---picked using validation data (1/4 test set sizes) as described in Section
\ref{sec:performancemetric} and (exact sizes in Appendix \ref{tab:train_test_sets})---and a rule-of-thumbs choice (Default). 
 
\begin{itemize}[leftmargin=0.2in]
    \item One-Class SVM (OCSVM) \cite{scholkopf2000support} (implemented by
        \cite{zhao2019pyod}). We choose a Gaussian kernel of the form $K(x,
        x') \propto \exp(-\|x-x'\|^2/2\sigma^2)$, and pick $\sigma$ as
        follows. We consider the quantiles $[0.1, 0.2, \ldots, 0.9]  \cup
        \{0.95\}$ on increasing distances between pairs of points in the
        training sample.\\ {\bf Optimal:} The parameter $\sigma$ is then picked as
        that quantile which yields the best AUC.\\ {\bf Default:} $\sigma$ is
        picked as the 0.3 quantile of inter-point distances. 

\item Isolation Forest (IF) \cite{liu2012isolation} (implemented by \cite{zhao2019pyod}). We
        pick the number $k$ of trees in the range 30 to 300 with increments of
        10. \\
    {\bf Optimal:} $k$ is chosen as the value in the range maximizing AUC. \\
    {\bf Default:} $k$ is 100, which is the same as the default value in pyod.

\item AutoEncoder (AE) \cite{zhou2017anomaly} (from PyTorch \cite{NEURIPS2019_9015} using LeakyRelu for activation functions). For $X\in \mathbb{R}^d$, we use the following architecture with 5 layers, determined by a tuning parameter $\text{dim}$: one input layer of size $d$ (number of \emph{neurons}), followed by a hidden layer of size $\text{h-dim} \doteq \text{min}\{(d-1), \lceil 2\cdot \text{dim}\rceil \}$, a latent layer of size $\text{dim}$, a subsequent hidden layer of size $\text{h-dim}$, and finally an output layer of size $d$. {We pick $\text{dim}$ in a range 
    $\{\lceil 1 + i\cdot (d -2)/9 \rceil \}_{i = 0}^9$
    of (up to 10) values (approximately evenly spaced) between $1$ and $d-1$}.
        \\
    {\bf Optimal:} $\text{dim}$ is chosen as the value in the range maximizing
        AUC. \\ {\bf Default:} $\dim$ is $\lceil d/2\rceil $. 
      
    \item Kernel Density Estimation (KDE) \cite{latecki2007outlier, scott2005multidimensional}. We use a Gaussian kernel, of the same form as OCSVM above, and pick the \emph{bandwidth} $\sigma$ exactly the same way for both {\bf Optimal} and {\bf Default}. 

    \item Gaussian Mixture Model (GMM) \cite{Bishop2006Pattern}. We choose the number of Gaussian components $k$ on a range 
    $[1, 2, 5, 10, 15,$\\$ 20, 25, 30, 35, 40]$. \\
    {\bf Optimal:} $k$ is chosen as the value in the range maximizing AUC. \\
    {\bf Default:} $k$ is obtained as the number of high-density regions (or modes) in the training data, as obtained by the \texttt{quickshift++} procedure of \cite{jiang2018quickshift++}.

\item Principal Component Analysis (PCA) \cite{aggarwal2015outlier}. For $X \in \mathbb{R}^d$, i.e., having $d$ features, we pick the projection dimension $\text{dim}$ in a range. We pick $\text{dim}$ in a range 
    $\{\lceil 1 + i\cdot (d -2)/9 \rceil \}_{i = 0}^9$
    of (up to 10) values (approximately evenly spaced) between $1$ and
        $d-1$.\\
    {\bf Optimal:} $\text{dim}$ is chosen as the value in the range maximizing
        AUC. \\
    {\bf Default:} $\text{dim}$ is estimated by maximum likelihood (MLE) in scikit-learn. 
    
\end{itemize}

%% file: results-table.tex
\begin{table}[t!] \centering
\begin{scriptsize}
\begin{tabulary}{0.99\textwidth}{|C|C|C|>{\columncolor[gray]{0.9}}C|C|>{\columncolor[gray]{0.9}}C|}
% \begin{tabulary}{0.99\textwidth}{\columncolor[gray]{0.8} c c c c c c c c c}
\toprule Detector & Dataset & \Cell{STATS} & \Cell{SIZE} & \Cell{IAT}  & \Cell{SAMP-\\NUM}  \\ 
\midrule
\multirow{7}{*}{OCSVM} &UNB(PC1) & 0.48 & 0.77 & 0.81 & 0.76   \\
&UNB(PC4) & 0.55 & 0.73 & 0.86 &  0.75  \\ 
\cmidrule{2-6}
&CTU & 0.64 & 0.77 & 0.76 & 0.91\\ 
\cmidrule{2-6}
&MAWI & 0.88 & 0.47 & 0.45 & 0.59\\ 
\cmidrule{2-6}
&TV\&RT & 1.00 & 1.00 & 0.96 & 0.93 \\ 
\cmidrule{2-6}
&SFrig & 0.98 & 0.89 & 0.82  & 0.96  \\ 
&BSTch & 0.95 & 0.98 & 0.97 & 0.95 \\ 
\midrule
\multirow{7}{*}{IF} &UNB(PC1) & 0.49 & 0.52 & 0.63  & 0.77  \\ 
&UNB(PC4) & 0.60 & 0.47 & 0.70 & 0.81 \\ 
\cmidrule{2-6}
&CTU & 0.78 & 0.79 & 0.86  & 0.89 \\ 
\cmidrule{2-6}
&MAWI & 0.86 & 0.69 & 0.42  & 0.62 \\ 
\cmidrule{2-6}
&TV\&RT & 0.95 & 0.98 & 0.90 & 0.75 \\ 
\cmidrule{2-6}
&SFrig & 0.96 & 0.95 & 0.59  & 0.94 \\ 
&BSTch & 0.95 & 0.98 & 0.94  & 0.98   \\ 
\midrule
\multirow{7}{*}{AE} &UNB(PC1) & 0.68 & 0.63 & 0.70  & 0.83  \\ 
&UNB(PC4) & 0.79 & 0.68 & 0.71 & 0.84 \\ 
\cmidrule{2-6}
&CTU & 0.49 & 0.83 & 0.93 &  0.91 \\ 
\cmidrule{2-6}
&MAWI & 0.49 & 0.62 & 0.46  & 0.60 \\ 
\cmidrule{2-6}
&TV\&RT & 1.00 & 1.00 & 0.97  & 0.85 \\ 
\cmidrule{2-6}
&SFrig & 0.91 & 0.92 & 0.73  & 0.97  \\ 
&BSTch & 0.96 & 0.94 & 0.97 & 0.97   \\ 
\midrule
\multirow{7}{*}{KDE} &UNB(PC1) & 0.31 & 0.77 & 0.83  & 0.72   \\ 
&UNB(PC4) & 0.52 & 0.71 & 0.86 & 0.75 \\ 
\cmidrule{2-6}
&CTU & 0.76 & 0.83 & 0.77  & 0.92 \\ 
\cmidrule{2-6}
&MAWI & 0.86 & 0.55 & 0.47 & 0.59 \\ 
\cmidrule{2-6}
&TV\&RT & 1.00 & 1.00 & 0.97 & 0.89\\ 
\cmidrule{2-6}
&SFrig & 0.97 & 0.90 & 0.80 & 0.95\\ 
&BSTch & 0.94 & 0.98 & 0.95 & 0.95\\
\midrule
\multirow{7}{*}{GMM} & UNB(PC1) & 0.56 & 0.61 & 0.75 & 0.80 \\ 
& UNB(PC4) & 0.87 & 0.71 & 0.65 & 0.81 \\ 
\cmidrule{2-6}
& CTU & 0.69 & 0.70 & 0.90 & 0.90 \\ 
\cmidrule{2-6}
& MAWI & 0.87 & 0.67 & 0.49 & 0.63 \\ 
\cmidrule{2-6}
& TV\&RT & 1.00 & 1.00 & 0.98 & 0.93 \\ 
\cmidrule{2-6}
& SFrig & 0.98 & 0.97 & 0.88 & 0.96 \\ 
& BSTch & 0.99 & 0.98 & 0.95 & 0.97 \\ 
\midrule
\multirow{7}{*}{PCA} & UNB(PC1) & 0.74 & 0.27 & 0.41 & 0.77 \\ 
& UNB(PC4) & 0.72 & 0.60 & 0.74 & 0.75 \\ 
\cmidrule{2-6}
& CTU & 0.81 & 0.74 & 0.75 & 0.89 \\ 
\cmidrule{2-6}
& MAWI & 0.70 & 0.45 & 0.40 & 0.59 \\ 
\cmidrule{2-6}
& TV\&RT & 0.99 & 1.00 & 0.97 & 0.65 \\ 
\cmidrule{2-6}
& SFrig & 0.80 & 0.91 & 0.71 & 0.95 \\ 
& BSTch & 0.97 & 0.08 & 0.92 & 0.95 \\ 
\bottomrule \end{tabulary} 
\end{scriptsize}
\caption{AUCs for 6 ML approaches on basic feature representations.}
    \label{tab:OCSVM_IF_AE_and_KDE_with_best_parameters}
\end{table}

%% file: results.tex
\section{Results} \label{sec:results}

In this section, we explore how data representation affects the performance of
different novelty detection algorithms.  After presenting baseline model
accuracy across a range of models for a variety of novelty detection problems,
we investigate the effects of various feature representations on model
accuracy.  We first then investigate whether and how packet sizes and packet
header information ultimately affect novelty detection, for different types of
novelty, across detection approaches.  We then explore the effect of Fourier
domain representations.

\subsection{Baseline Model Accuracy}

In order to understand how different data representations affect model
accuracy, we first need to understand baseline accuracy for each model, 
scenario, and a baseline set of features. 
Because we are interested in the
relative effects of additional features and transformations, including packet
sizes, headers, and Fourier representations, we first present baseline levels
of accuracy across models without these features.
Table~\ref{tab:OCSVM_IF_AE_and_KDE_with_best_parameters} gives baselines for
comparisons. 
We are interested in the difference in AUC due to the presence or absence of
particular feature types with respect to a set of basic feature
representations, namely STATS, SIZE, IAT, SUMP-NUM, and SAMP-SIZE.  

\begin{figure*}[t!] \centering
    \includegraphics[width=0.95\textwidth]{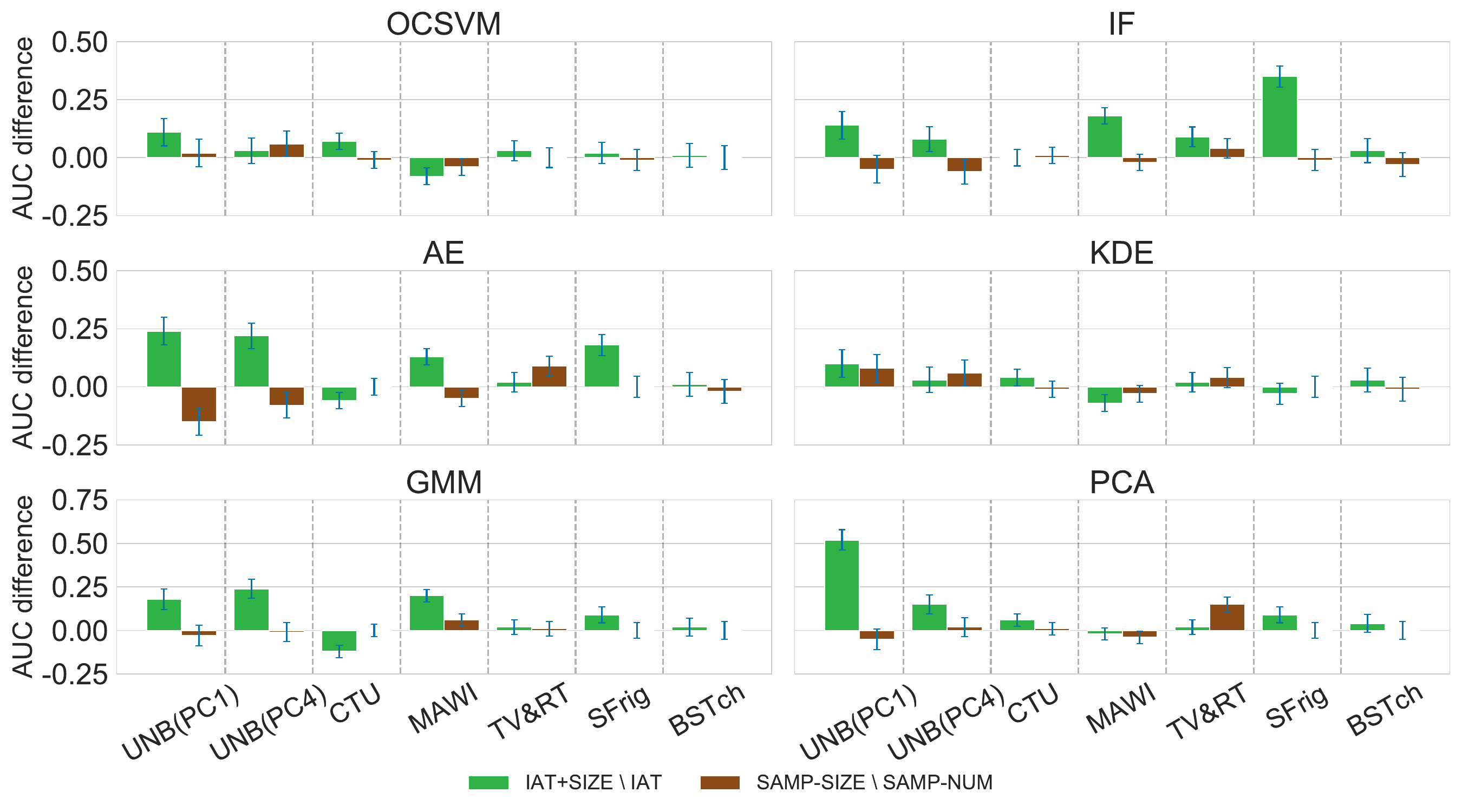}
\caption{Differences in AUC due to {\bf including vs. excluding packet size}
    information.}
\label{fig:diffinsize} \end{figure*}

Table \ref{tab:OCSVM_IF_AE_and_KDE_with_best_parameters} also includes
IAT+SIZE as a baseline for future comparisons). We will see that across ML
models, the effect of (1) packet size information, and (2) packet header
information tend to be significant overall.  (3) Fourier domain
representations have little advantage over raw time series (despite being
generally thought of as a better way to capture important trends in
timeseries)~\cite{oppenheim1999discrete}.
We focus on the most popular ML approaches disxcussed so far (OCSVM, IF, AE, KDE, PA, GMM) in the
main text, under {\bf OPT} tuning, while results under {\bf Default} tuning, and omitted datasets are given in the appendix. The
choices of procedures is meant to be representative of each major family of
approach as described in Section \ref{sec:ml_procedures}.
% and PCA are similar in that they both aim for a lower-dimensional intrinsic
% representation of the data, and KDE and GMM are similar in that they both aim
% to estimate regions of high density of the data. 

\paragraph{Statistical significance.} Throughout this section, we often
display error bars of length $1/\sqrt{n}$ ($n$ denoting test sample sizes as
given in Table \ref{tab:train_test_sets}), which is a ballpark for standard
deviation in AUC computed on a test sample of size $n$. Training data sizes are for the most part around 5000 (see again Table  \ref{tab:train_test_sets}) besides for a few cases where they are smaller to allow large enough test size $n$ (at least a few hundred points to ensure statistical significance).  

\subsection{Effects of Data Representations}

Given an understanding of baseline performance, we now explore the effects of
various representations of network traffic on model accuracy across the range
of novelty detection models and scenarios we described earlier.

\subsubsection{Packet Size Information}\label{sec:size}

Table~\ref{tab:OCSVM_IF_AE_and_KDE_with_best_parameters} shows that packet size
information immediately appears to be important based on the often higher
AUCs achieved under STATS (which primarily compiles statistics on packet sizes
in a flow) and SIZE (time series of packet sizes) to those achieved under basic representations such as IAT and
SAMP-NUM (devoid of packet size information). This is most evident, e.g., under 
for the MAWI and SFrig datasets. 
%Interestingly, SIZE, a time
%series of packet sizes in a flow, does not perform as well as STATS does over
%IAT, and SAMP-NUM. 

As it turns out, adding packet size information to IAT or SAMP-NUM
representations improves on such representations alone.  Namely, in the case
of IAT, size information is added in by concatenating IAT and SIZE vectors,
which we denote IAT+SIZE. For SAMP-NUM, which is a time-series of \emph{number of
packets} in small fixed time intervals $\delta t$, size information is
integrated in by instead compiling \emph{total packet size} in fixed time
intervals $\delta t$.  %{\color{brown} The reported AUCs are for the best
%choice of $\delta t$, chosen for each approach as detailed in Section
%\ref{sec:data_representations}.}

Figure \ref{fig:diffinsize} presents the differences in AUC obtained using
packet size information, minus those obtained on corresponding representations
without size information. We observe significant positive differences for the features IAT+SIZE, across
ML procedures on most datasets---which we recall, correspond to a range of
novelty detection problems, whether malicious attacks, or novel devices or
activity. Moreover, in most cases where size information does not yield improvements, it nonetheless does not hurt performance (changes are under the significance level shown by error bars), except in the case of dataset UNB(PC1) under SAMP when
employing AE. This is however a case where packet size information remains
useful with respect to IAT, just not with respect to sampling-based features SAMP. As such, it
appears that for the \emph{sampling} based features SAMP-SIZE and SAMP-NUM,
\emph{number of packets} or \emph{packet sizes} in fixed time intervals are
equally predictive in general. One reason for this observed effect is likely
because packet sizes are predictable and are often even uniform in size,
particularly for long-running and high-volume flows, where packet size is
limited by the maximum transmission unit (MTU).

% However, even in this case, we observe (see Table
% \ref{tab:OCSVM_IF_AE_and_KDE_with_best_parameters}) that the concatenation
% of IAT+SIZE yields {\em worse} AUC than IAT or SIZE alone. 

%{\color{blue}The exact reason is unclear and could be due to
%\emph{overfitting}, as the concatenation increases feature dimension in ways
%that might affect \emph{nonparametric} models such as OCSVM and KDE given the
%relatively modest training size of MAWI (5000 datapoints).}

% The most significant positive differences are observed for IF, AE, and PCA
% in the case of IAT, while no significant differences are observed for
% sampling-based representations, such as SAMP-SIZE vs. SAMP-NUM. 

% Interestingly, for fixed datasets, trends are generally consistent across ML
% methods, suggesting that whether size information helps depends most on
% devices' characteristics, rather than the ML method employed or the type of
% detection problem. 
\begin{figure*}[t] \centering
    \includegraphics[width=0.95\textwidth]{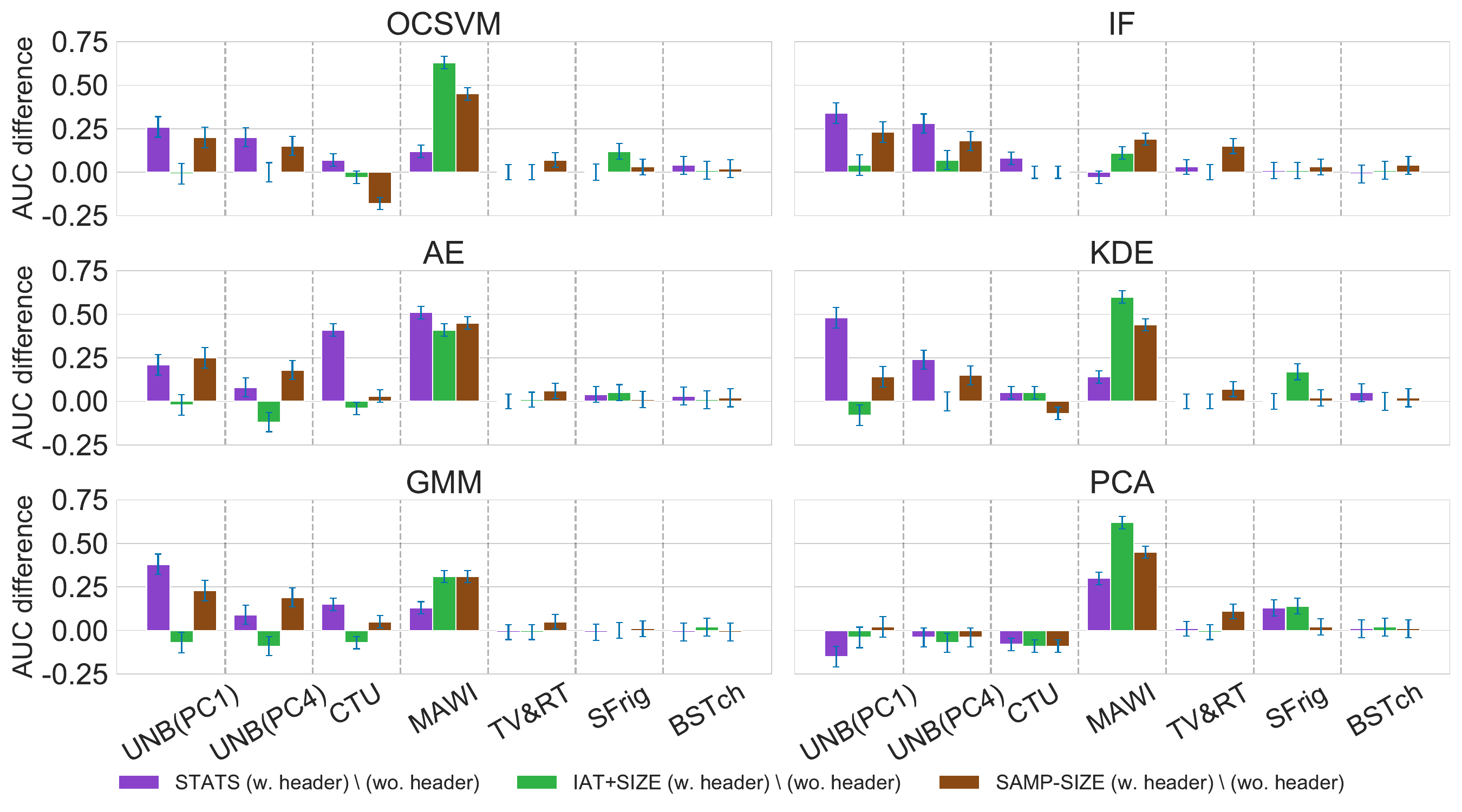}
\caption{Differences in AUC due to {\bf including vs. excluding packet header}
information.} \label{fig:diffpacketheader} \end{figure*}

In general, the inclusion of packet sizes for novelty detection makes sense,
as the packet size can provide some indication about the nature of the
underlying activity. For example, for the case of the UNB dataset, attack
traffic may be unique in size relative to other types of activities. In the
case of the private IoT datasets (TV\&RT, SFrig, BSTch), new activities may be
less likely to produce individual packets of differing sizes for the same set
of activities. On the other hand, a new activity could produce flows with
packets of different sizes than previously observed from the existing set of
activities, particularly if that activity is low-volume, as is common with
some smart home activities.

\subsubsection{Packet Flags and Headers} \label{sec:header} 

We now consider the effects of information in packet headers on novelty
detection, focusing on the information available in various TCP flags. We
focus on the following eight flags in each packet: FIN, SYN, RST, PSH, ACK,
URG, ECE, and CWR, and TTL. We do not include IP address or port information
in this part of our analysis.  Given the already established benefits (or at
least invariance in AUC) of including packet size information, we now take
representations such as STATS, IAT+SIZE and SAMP-SIZE as baselines to which we
concatenate packet header information (for each packet in a flow) and record
differences in performance. 

Figure \ref{fig:diffpacketheader} presents the differences in AUC obtained
with packet header information minus that obtained without header information.
In general we observe no \emph{statistically} significant degradation in
performance in adding header information, except in the rare case of (i) CTU,
using SAMP-SIZE features under OCSVM, (ii) to a lesser extent, UNB(PC4) using
IAT+SIZE under AE, and (iii) UNB(PC1) using STATS under PCA.

In terms of performance improvements due to header information, similar trends
are observed across ML approaches. As might be expected, {we observe
significant improvements on those datasets corresponding to (1) detecting
malicious activity (UNB, CTU)}---where infected traffic might be rerouted to
new destinations, thereby resulting in traffic with different distributions of
packet TTL values corresponding to the different, new destinations---or (2)
detecting novel devices (CTU, MAWI)---where header information such as certain
TCP options are often specific to a particular device or operating system.
Interestingly, for less obvious reasons (discussed below), we observe
improvements in some cases for \emph{novel activity detection}, namely for the
SFrig dataset under   OCSVM, KDE, and PCA.

Although TV\&RT seem to be an exception to (2), the general lack of
improvement is simply due to the fact that AUC's for the baseline
representations were already near perfect (close to 1) across ML approaches
(see Table \ref{tab:OCSVM_IF_AE_and_KDE_with_best_parameters}).  This happens
to also be the case (i.e., near perfect AUC's) for those baseline
representations where we observe little difference in AUC for the two
detection problems (1) and (2). The most improvement across ML methods is
observed for MAWI where the baseline representations yielded poor AUC's to
start with (Table \ref{tab:OCSVM_IF_AE_and_KDE_with_best_parameters}). 

\begin{figure*}[t!] \centering
\includegraphics[width=0.95\textwidth]{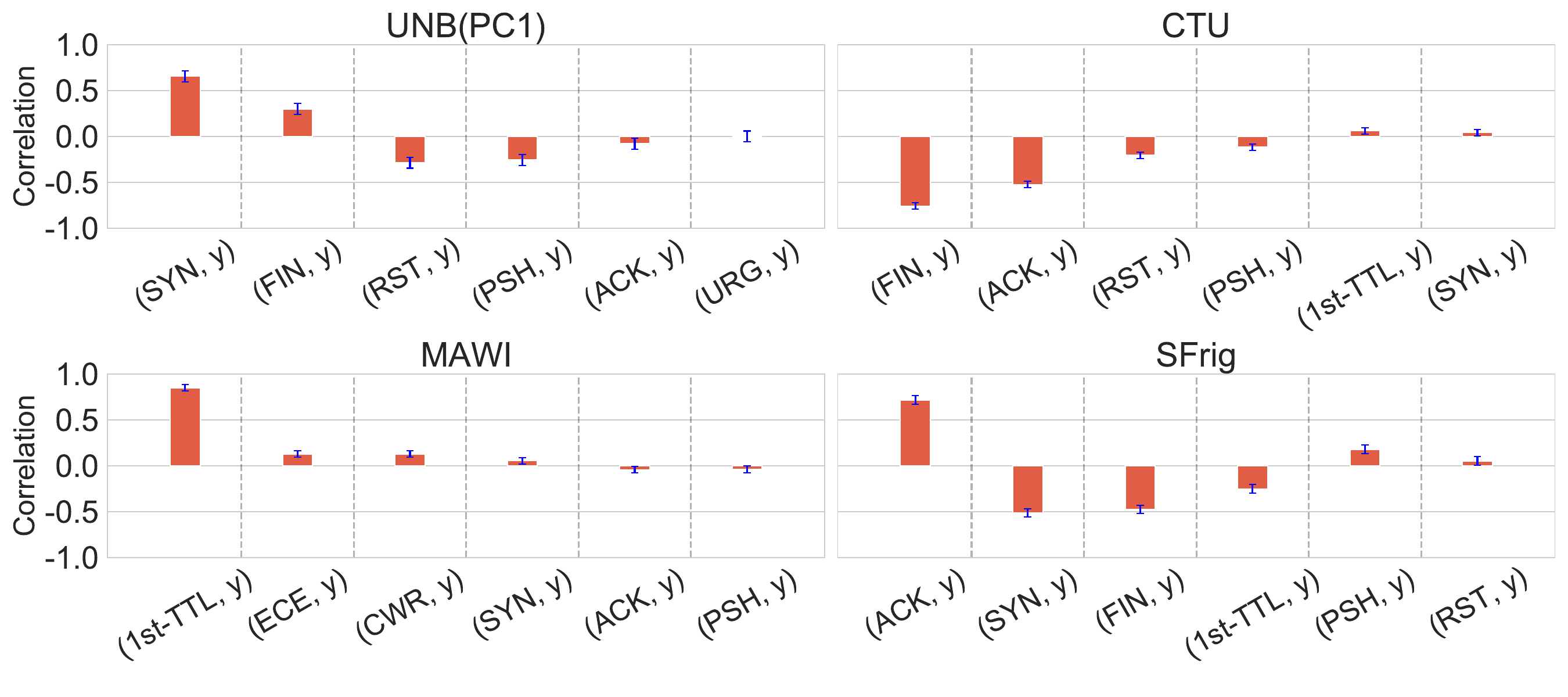} \caption{Top 6
correlation between each value in packet headers and ground truth ($y$).}
\label{fig:correlation} \end{figure*}

\paragraph{Relative importance of specific header flags:}
Figure~\ref{fig:correlation} shows that certain packet header fields can be
particularly useful for novelty detection, depending on the anomaly; these
results are evident in the results. One of the more important findings from
taking a closer look at the correlations between features and ground truth is
that the most important packet header values will {\em depend on the novelty
detection problem}. We explore this in more detail below.

\paragraph{Attack detection:} In the case of the UNB dataset, we see that the SYN
flag can be very useful in identifying novelty events related to security, as
the SYN flag corresponds to a TCP handshake that occurs at the beginning of
the connection and is commonly related to certain types of denial of service
attacks, such as a SYN flood. Other packet header information that exhibits
correlation include other flags that are commonly associated with denial of
service attacks, including the RST and FIN flags, which victim hosts may send
in response to a SYN. Traffic flows may also see a novel distribution of
packets with ACK flags, given that packets with ACK flags are sent in response
to SYN packets, which are sent either to initiate a new flow or as part of an
attack (e.g., a SYN flood attack).  Such a situation is the case with CTU
(novel infected device) where we see high correlation for FIN, ACK, and RST
flags.  In all of these cases, an increase in packets that include these flags
could likely correspond with a volumetric denial of service attack, such as a
SYN flood.  
\begin{figure*}[t] \centering
    \includegraphics[width=0.95\textwidth]{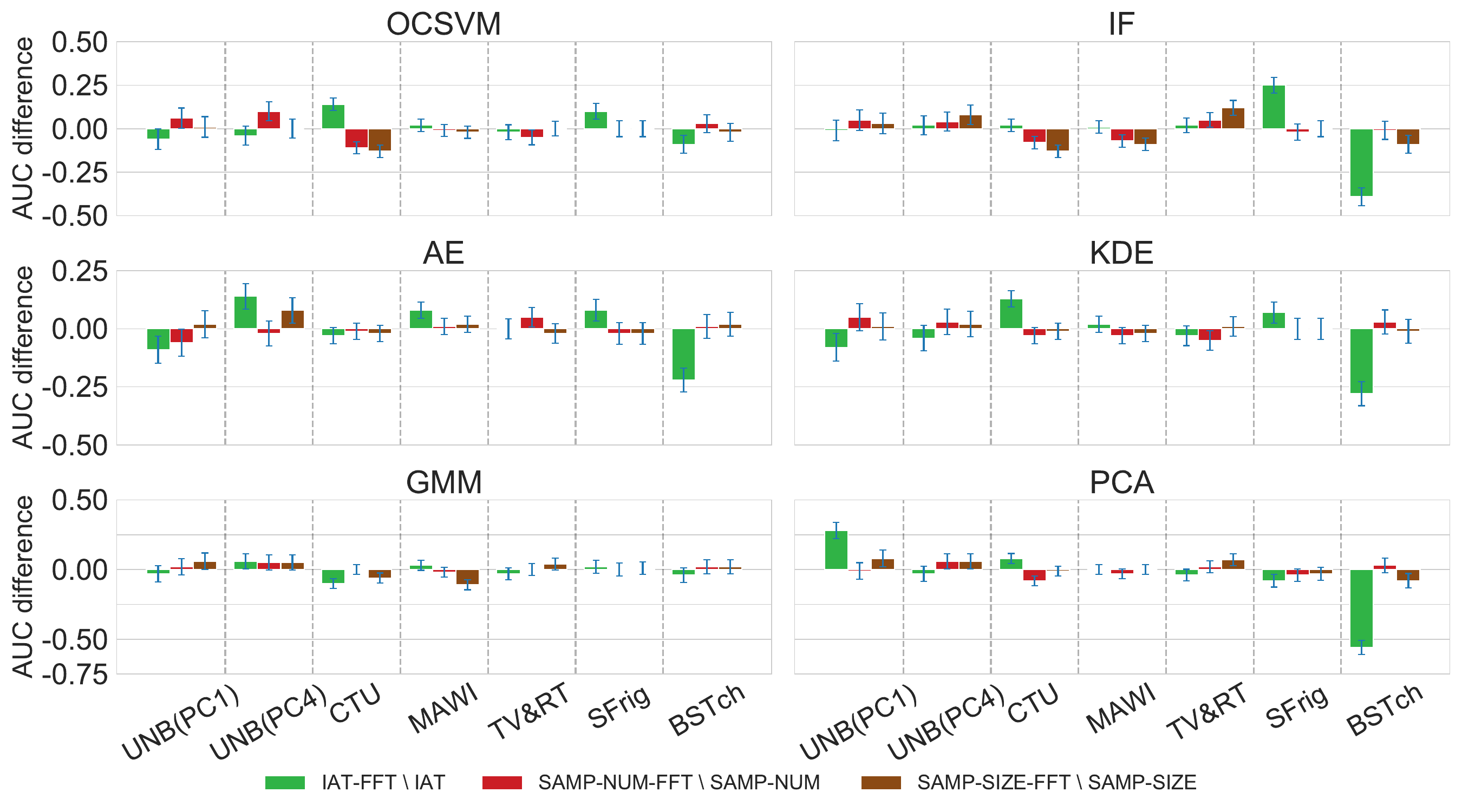}
    \caption{Difference in AUC for {\bf FFT vs. raw time series}
    representations.}
\label{fig:diffinFFT} \end{figure*}

\paragraph{Novel device detection:} The MAWI dataset, where the novelty detection
problem involves detecting a novel device, finds the TTL field in a packet
header to be significant. This result also follows intuition, as the TTL value
is often a coarse indicator of network topology (i.e., how many network-level
``hops'' a device is from a particular destination endpoint) and naturally two
different devices on the network may be located at different network
attachment points and thus be different distances away from common network
locations. Such topological differences would be apparent in the TTL field,
and a new device attaching at a different location on the network could appear
as a sudden influx of network traffic that bears a different distribution in
TTL values.  In particular, sudden deviations in distributions of TTL values
in the dataset may thus reflect the injection of traffic from new devices, or
possibly attackers. In the CTU and Lab IoT datasets, the TTL field is
comparatively less important because the devices in the dataset were connected
on the same local area network, as opposed to different places in the wide
area. In such a scenario, the connected devices are a single IP hop away from
the monitoring location and thus the TTL values will be the same for all
devices. The importance of the SYN field varies across datasets and scenarios,
most likely because in some cases the introduction of a new device may involve
a change in the distribution of new TCP flows (where the start of each flow is
characterized by a SYN packet), whereas in other cases, changes in the number
of new flows may not be a useful feature for detecting a novel device.

\paragraph{Novel activity detection:} If someone begins to interact with a
connected IoT device in new ways---triggering new types of activities---the
traffic itself may bear differences from previously seen traffic. Differences
in the traffic itself may arise because activities themselves may bear unique
signatures in the underlying network traffic. For example, the amount and
nature of traffic generated by opening and closing a refrigerator door in
SFrig will generate different types of traffic than the kind of traffic that
is generated by activities that do not involve human interaction (e.g., a
software update).

These fundamental differences may appear in different ways, such as
differences in traffic volume or timing, and in some cases they may appear in
the TCP header flags themselves. In the case of SFrig, we see that the ACK
flag has a high correlation with ground truth, indicating that changes in the
prevalence of ACKs (i.e., packet acknowledgments) may represent changes in
underlying activities. This characteristic likely results from the underlying
behavior of TCP, the Internet's transport protocol, and how it handles packet
acknowledgments for flows of different sizes and timings. For example, for
large traffic flows, TCP will sometimes optimize its acknowledgment behavior,
using techniques such as {\em delayed ACK} to improve network performance.
Highly interactive behavior and the traffic it generates may exhibit very
different qualities than automated traffic (e.g., a software update involving
the transfer of a large amount of data) and thus different behavior in ACK
traffic.  The distribution of ACK fields in the traffic may reflect higher
traffic volumes during periods of novel activity.  Changes in distributions or
frequencies of SYN and FIN packets in the trace can also indicate changes in
the number of overall total flows, reflecting novel activities that generate
completely new flows or simply change the distribution of active flows.
Finally, as stated earlier, the lack of improvement for Sfrig using SAMP-SIZE
or STATS features is due to the already high performances achieved under these
features on most methods. The same reason applies for the BSTch dataset, where
detection is nearly solved under any feature representation.

\subsubsection{Fourier Domain Representation}\label{sec:fft}

We compare Fourier domain representations (obtained by \emph{Fast Fourier
    Transforms} (FFT), where we retain as many FFT components as the number of
    time points) to raw time series representations of each flow, namely IAT,
    SAMP-NUM, and SAMP-SIZE. Figure \ref{fig:diffinFFT} presents the difference in
    AUC, i.e., the AUC obtained under FFT minus that obtained under the
    corresponding raw time series representation. 

A similar trend is observed across ML procedures: apart for a few device
    datasets (e.g., positive effects in SFrig, and in UNB (PC1, PC4) --- under AE and PCA --- and negative effect in BSTch
    with respect to IAT), FFT genrally makes no significant difference in the achieved AUC over
    raw time series. In other words, the various ML procedures seem able to
    extract much of the same information already from the raw time series,
as their \emph{internal representations} of the data already account
    for major trends in the times series and thus do away with the need to
    preprocess through FFT.

% This part reports the results with \IAT, \IATFFT, \STATS, \SAMPnum, and
    % \SAMPnumFFT, and for SAMP-based results, we show the best AUC with the
    % corresponding fraction ($q_s$) used to obtain the sampling rate from
    % flow durations, and the minimum AUC.

%% file: conclusion.tex
\section{Conclusion}\label{sec:conclusion}

The performance of novelty detection in computer networking requires careful
decisions about how to best represent traffic data, e.g, which predictive
features to extract, which time windows to use, whether to work in time domain
or Fourier domain, etc. As we have seen, these choices can significantly
affect detection accuracy and depend specifically on the the types of novelty
we seek to detect and the machine learning model being used.

Unfortunately, despite much past work on anomaly detection in networking, we
have no general guiding principles towards predictive yet succinct traffic
data representation. Indeed, past work in this area has focused on solutions
to individual problems, with application of a specific model (e.g., PCA) to a
particular representation (e.g., IPFIX/NetFlow).  Without a general framework,
researchers and practitioners must apply heuristics to select models, model
parameters and representations, a problem which is further exacerbated by the
general lack of automation of these common preprocessing steps. As a result,
each problem in networking that seeks to address anomaly detection (or, more
generally, novelty detection) must re-invent the wheel, developing bespoke
techniques for each step and generating pipelines that are difficult to
reproduce.

To address these difficulties, we have developed and released an open-source
library and command-line tool, {\tt netml}, that extracts common features
from network packet captures. The tool is configurable to allow for specific
application choices of features, yet simply to use and upgrade to incorporate state of the
art models and representations. For example, we have recently extended the
library to develop and test a fast, online OCSVM specifically designed for
network anomaly detection settings. It is our hope and expectation that these
types of tools will facilitate more such research across the network
measurement and machine learning communities.

%% file: appendix.tex
\appendix
\onecolumn

% \renewcommand{\thefigure}{A\arabic{figure}}
% \setcounter{figure}{0}
% \section*{Appendices}
% \appendices

\counterwithin{figure}{section} 
\counterwithin{table}{section}

\section{Further Information on Datasets}
\begin{table*}[hbp!]
\centering
\caption{Data representation dimensions for each dataset.}
\label{tab:feature_dimensions}
% \begin{tabular}{|p{1.3cm}|>{\centering\arraybackslash}p{1.5cm}|>{\centering\arraybackslash}p{1.5cm}|>{\centering\arraybackslash}p{1.cm}|>{\centering\arraybackslash}p{2.5cm}|} 
\begin{tabular}{|c|c|c|c|c|}
\toprule
Device & \Cell{IAT,\\IAT-FFT} & \Cell{SIZE,\\SIZE-FFT} & \Cell{STATS} & \Cell{SAMP-NUM,\\SAMP-NUM-FFT,\\SAMP-SIZE} \\ 
\midrule
\multirow{5}{*}\ PC1 & 14&15&10&14 \\ 
\ PC2 & 16&17&10&16 \\ 
\ PC3 & 16&17&10&16 \\ 
\ PC4 & 19&20&10&19 \\ 
\ PC5 & 15&16&10&15 \\
\midrule
\ 2Rsps & 12&13&10&12\\ 
\midrule
\ 2PCs & 35&36&10&35\\ 
\midrule
\multirow{5}{*}\ TV\&RT & 12&13&10&12 \\ 
\ GHom & 19&20&10&19 \\ 
\ SCam & 6&7&10&6 \\ 
\ SFrig & 30&31&10&30 \\ 
\ BSTch & 148&149&10&148  \\ 
\bottomrule
\end{tabular}
\end{table*}

\begin{table*}[hbp!]
\centering
\caption{Train, Validation, and Test set, in which 'N' represents normal and 'A' represents novelty (e.g., attack, novel device, and novel activity). Validation set is only used for picking the best parameters for each algorithm and its size is a quarter of the size of each test set.}
\label{tab:train_test_sets}
\begin{tabular}{|c|c|l|l|l|}
\toprule
Reference & Devices & Train set  & Validation set  & Test set  \\
\midrule
\multirow{5}{*}{UNB IDS } & PC1 & N: 5000 & N: 35, A: 35 & N: 144, A: 144 \\
 & PC2 & N: 5000 & N: 51, A: 51 & N: 206, A: 206 \\
 & PC3 & N: 5000 & N: 44, A: 44 & N: 177, A: 177 \\
 & PC4 & N: 5000 & N: 42, A: 42 & N: 168, A: 167 \\
 & PC5 & N: 5000 & N: 72, A: 72 & N: 288, A: 288 \\
\midrule
% HCR IoT  & SKT camera & N: 5021 & N: 400 Outlier: 400 \\
% \midrule
CTU IoT  & 2Rsps & N: 5000 & N: 100, A: 100  & N: 400, A: 400 \\
\midrule
MAWI  & 2PCs & N: 5000 & N: 100, A: 100 & N: 400, A: 400 \\
\midrule
\multirow{5}{*}{Lab IoT } & TV\&RT & N: 4636 & N: 69, A: 69 & N: 277, A: 277 \\
 & GHom & N: 5000 & N: 100, A: 100 & N: 400, A: 400 \\
 & SCam & N: 5000 & N: 40, A: 40 & N: 161, A: 161 \\
 & SFrig & N: 2952 & N: 59, A: 59 & N: 237, A: 237 \\
 & BSTch & N: 997 & N: 47, A: 47 & N: 189, A: 189  \\
\bottomrule
\end{tabular}
\end{table*}

\newpage
% \begin{footnotesize}
\section{Best Parameters: additional  datasets and Procedures}
\subsection{Baseline Results with Best Parameters}

% \twocolumn
\begin{table}[htp!]
\small
\centering 
\caption{AUCs for 4 ML approaches on basic feature representations with best parameters.}
\label{tab:OCSVM_IF_AE_KDE_GMM_PCA_with_best_parameters} 
\begin{tabular}{|c|c|g|c|g|c|}
% \centering 
% \endfirsthead
% \endhead
\toprule
%  Detector & Dataset & STATS & SIZE & IAT & IAT+SIZE & SAMP-NUM & SAMP-SIZE  \\ 
Detector & Dataset & \Cell{STATS} & \Cell{SIZE} & \Cell{IAT}
    & \Cell{SAMP-\\NUM}   \\
\midrule
\multirow{5}{*}{~\rule{0pt}{2.7ex}OCSVM} &UNB(PC2) & 0.59 & 0.78 & 0.79 & 0.80  \\ 
&UNB(PC3) & 0.64 & 0.86 & 0.81 & 0.79 \\ 
&UNB(PC5) & 0.49 & 0.74 & 0.80 & 0.80  \\ 
\cmidrule{2-6}
&GHom & 0.73 & 0.96 & 0.69 & 0.95 \\ 
% \cmidrule{2-6}
&SCam & 0.67 & 0.64 & 0.53 & 0.59\\ 
\midrule
\multirow{5}{*}{~\rule{0pt}{2.7ex}IF} &UNB(PC2) & 0.68 & 0.59 & 0.67 & 0.81   \\ 
&UNB(PC3) & 0.69 & 0.66 & 0.69  & 0.76 \\ 
&UNB(PC5) & 0.58 & 0.62 & 0.71 & 0.78 \\ 
\cmidrule{2-6}
&GHom & 0.88 & 0.78 & 0.42 & 0.96\\ 
% \cmidrule{2-6}
&SCam & 0.64 & 0.63 & 0.54  & 0.63 \\ 
\midrule
\multirow{5}{*}{~\rule{0pt}{2.7ex}AE} &UNB(PC2) & 0.72 & 0.62 & 0.73 & 0.83  \\ 
&UNB(PC3) & 0.78 & 0.60 & 0.70  & 0.80  \\ 
&UNB(PC5) & 0.69 & 0.68 & 0.68  & 0.91  \\ 
\cmidrule{2-6}
&GHom & 0.82 & 0.66 & 0.33 & 0.96 \\ 
&SCam & 0.67 & 0.46 & 0.54 & 0.61 \\ 
\midrule
\multirow{5}{*}{~\rule{0pt}{2.7ex}KDE} &UNB(PC2) & 0.56 & 0.77 & 0.79 & 0.79  \\ 
&UNB(PC3) & 0.62 & 0.86 & 0.81 & 0.79  \\ 
&UNB(PC5) & 0.43 & 0.71 & 0.80 & 0.79 \\ 
\cmidrule{2-6}
&GHom & 0.57 & 0.96 & 0.70 & 0.96 \\ 
&SCam & 0.68 & 0.63 & 0.47  & 0.56\\ 
\midrule
\multirow{5}{*}{~\rule{0pt}{2.7ex}GMM} & UNB(PC2) & 0.78 & 0.63 & 0.71 & 0.81 \\ 
& UNB(PC3) & 0.87 & 0.77 & 0.77 & 0.81 \\ 
& UNB(PC5) & 0.81 & 0.72 & 0.74 & 0.86 \\ 
\cmidrule{2-6}
& GHom & 0.97 & 0.96 & 0.91 & 0.96 \\ 
& SCam & 0.54 & 0.64 & 0.47 & 0.63 \\ 
\midrule
\multirow{5}{*}{~\rule{0pt}{2.7ex}PCA} & UNB(PC2) & 0.69 & 0.62 & 0.43 & 0.79 \\ 
& UNB(PC3) & 0.71 & 0.28 & 0.32 & 0.81 \\ 
& UNB(PC5) & 0.67 & 0.30 & 0.38 & 0.75 \\ 
\cmidrule{2-6}
& GHom & 0.72 & 0.40 & 0.05 & 0.96 \\ 
& SCam & 0.63 & 0.61 & 0.62 & 0.59 \\ 
\bottomrule
\end{tabular}
\end{table}

\newpage
\subsection{Effect of Fourier Domain Representation}
% \onecolumn
The figures below show the difference in AUC for the FFT vs. raw timeseries
representations with the best model parameters for each model. FFT
transformations have little effect on model accuracy.

\begin{figure}[H]
\centering
\includegraphics[width=0.95\textwidth ]{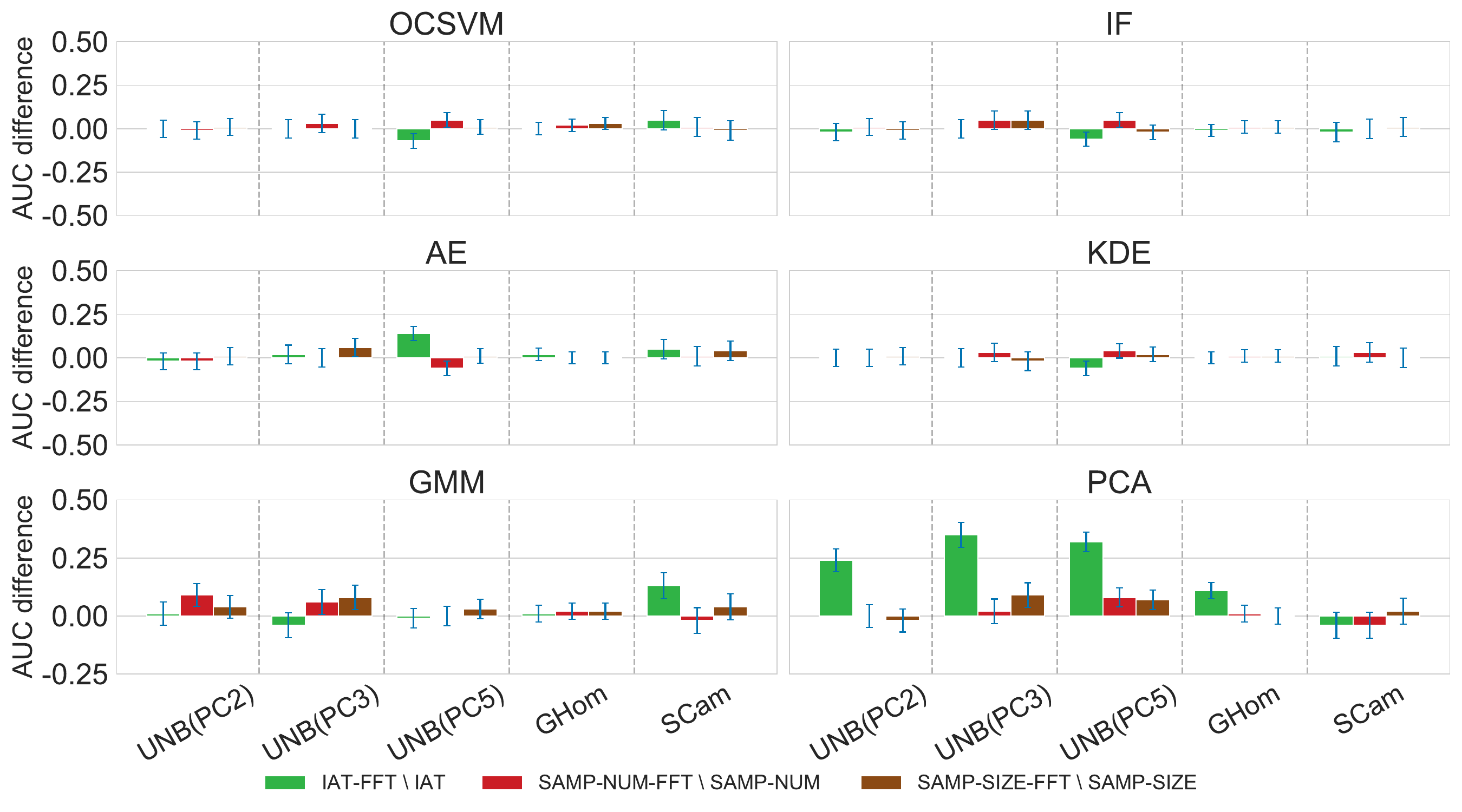}
\caption{Difference in AUC for {\bf FFT vs. raw time series} representations with best parameters.}
\label{fig:OCSVM_IF_AE_KDE_GMM_PCA-5-best-basic_representation}
\end{figure}

% \begin{figure}[H]
% \centering
% \includegraphics[width=0.95\textwidth]{GMM_PCA-12-best-basic_representation.pdf}
% \caption{Difference in AUC for {\bf FFT vs. raw time series} representations generated by GMM and PCA with best parameters.}
% \label{fig:GMM_PCA-12-best-basic_representation}
% \end{figure}

\newpage
\subsection{Effect of Packet Size Information}
\pagestyle{plain} % removes running headers

The figures below show the difference in AUC for including vs. excluding
packet size information for different models: IAT and SAMP. Including packet
size can significantly improve model accuracy in certain cases.

\begin{figure}[H]
\centering
\includegraphics[width=0.95\textwidth, height=9.5cm ]{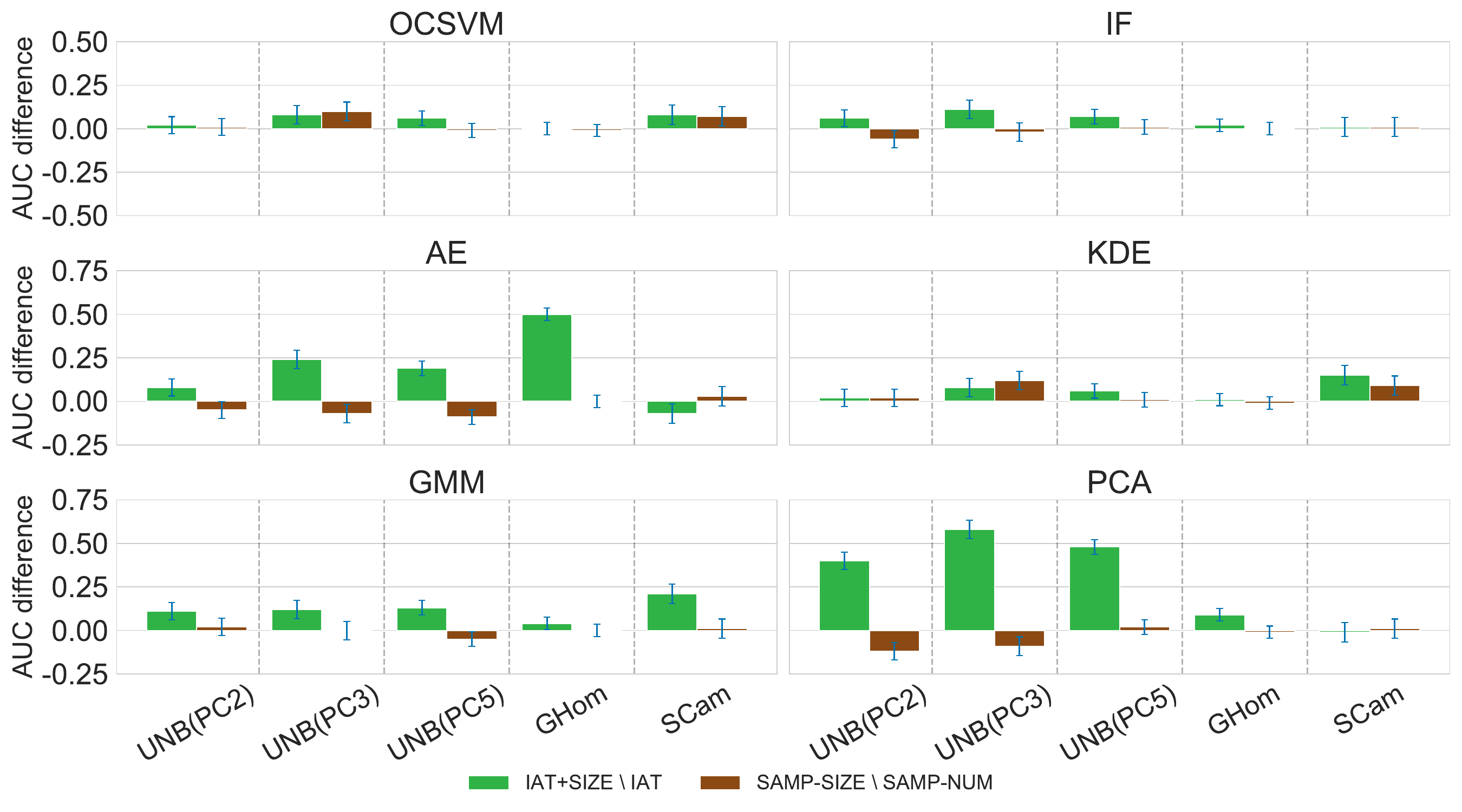}
\caption{Differences in AUC due to {\bf including vs. excluding packet size} information with best parameters.}
\label{fig:OCSVM_IF_AE_KDE_GMM_PCA-5-best-effect_size}
\end{figure}

% \begin{figure}[H]
% \centering
% \includegraphics[width=0.95\textwidth, height=8.5cm]{GMM_PCA-12-best-effect_size.pdf}
% \caption{Differences in AUC due to {\bf including vs. excluding packet size} information generated by GMM and PCA with best parameters.}
% \label{fig:GMM_PCA-12-best-effect_size}
% \end{figure}

\newpage
\subsection{Effect of Packet Header}
The figures below show the difference in AUC for including vs. excluding
packet header information (i.e., IP TTL, TCP flags) for different models:
STATS, IAT, and SAMP. Including packet
header information almost always improves model accuracy for these models.

\begin{figure}[H]
\centering
\includegraphics[width=0.95\textwidth,height=8.5cm ]{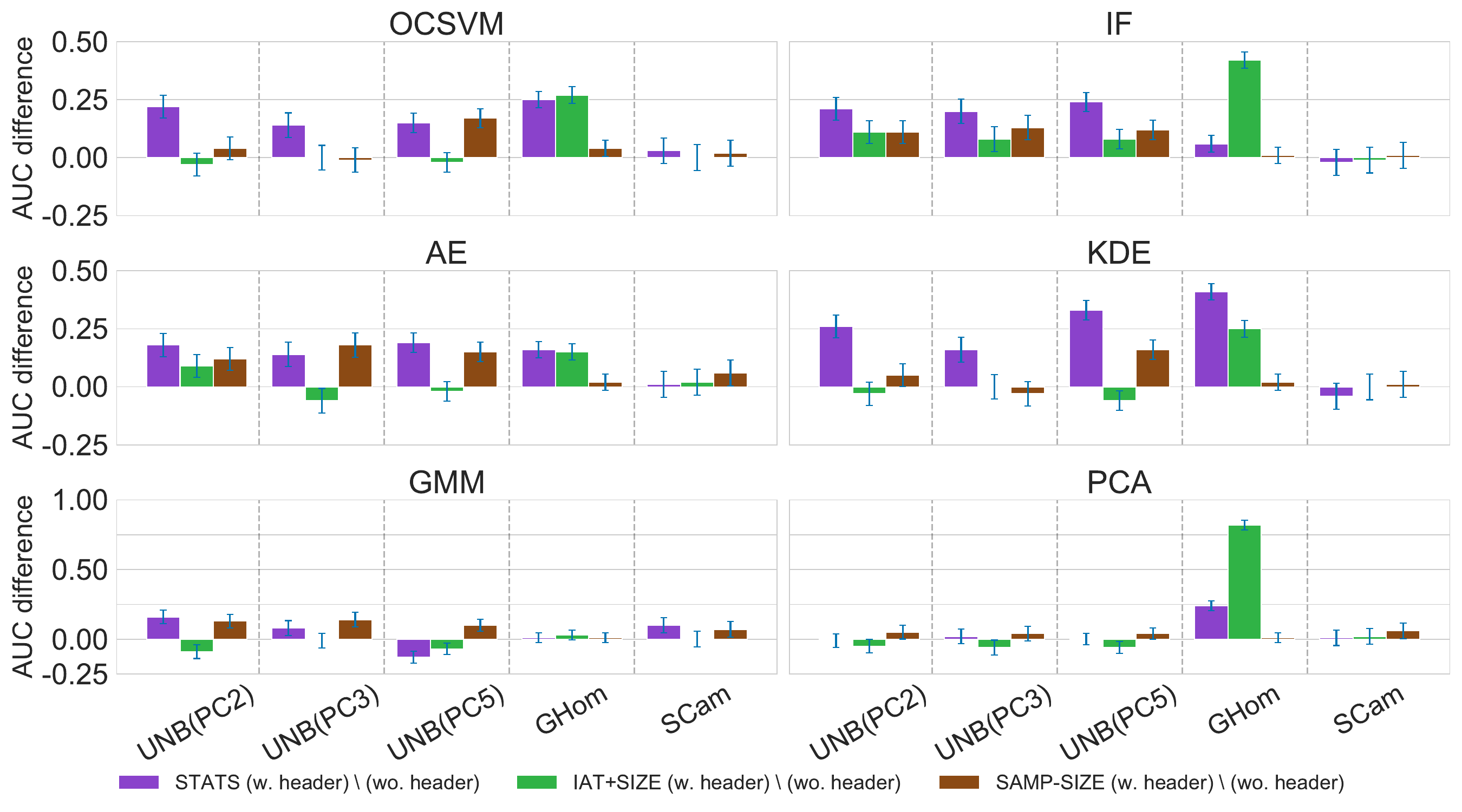}
\caption{Differences in AUC due to {\bf including vs. excluding packet header} information with best parameters.}
\label{fig:OCSVM_IF_AE_KDE_GMM_PCA-5-best-effect_header}
\end{figure}

% \begin{figure}[H]
% \centering
% \includegraphics[width=0.95\textwidth,height=8.5cm]{GMM_PCA-12-best-effect_header.pdf}
% \caption{Differences in AUC due to {\bf including vs. excluding packet header} information generated by GMM and PCA with best parameters.}
% \label{fig:GMM_PCA-12-best-effect_header}
% \end{figure}

\newpage
\section{Default Parameters: Additional Datasets and Procedures}

\subsection{Baseline Results with Default Parameters}
% \newpage
% \begin{scriptsize}
\begin{table*}[htbp!]
\centering 
\begin{threeparttable}
\centering
\caption{AUCs for 4 ML approaches on basic feature representations with default parameters.}
\label{tab:OCSVM_IF_AE_and_KDE_with_default_parameters} 
\begin{tabular}{|c|c|g|c|g|c|g|c|}
% \endfirsthead
% \endhead
\toprule
 Detector & Dataset & \Cell{STATS} & \Cell{SIZE} & \Cell{IAT} &  \Cell{SAMP-\\NUM} \\ 
\midrule
\multirow{12}{*}{~\rule{0pt}{2.7ex}OCSVM}  & UNB(PC1) & 0.48 & 0.42 & 0.72 & 0.76 \\ 
 & UNB(PC2) & 0.48 & 0.37 & 0.79 & 0.79 \\ 
 & UNB(PC3) & 0.54 & 0.80 & 0.82 & 0.78 \\ 
 & UNB(PC4) & 0.55 & 0.62 & 0.86 & 0.75 \\ 
 & UNB(PC5) & 0.49 & 0.64 & 0.74 & 0.77 \\ 
 \cmidrule{2-6}
 & CTU & 0.54 & 0.77 & 0.75 & 0.85 \\ 
 \cmidrule{2-6}
 & MAWI & 0.28 & 0.28 & 0.41 & 0.58 \\ 
 \cmidrule{2-6}
 & TV\&RT & 1.00 & 1.00 & 0.95 & 0.68 \\ 
 \cmidrule{2-6}
 & GHom & 0.73 & 0.96 & 0.07 & 0.95 \\ 
 & SCam & 0.68 & 0.46 & 0.53 & 0.56 \\ 
 & SFrig & 0.88 & 0.90 & 0.77 & 0.95 \\ 
 & BSTch & 0.97 & 0.97 & 0.93 & 0.92 \\ 
\midrule
\multirow{12}{*}{~\rule{0pt}{2.7ex}IF} & UNB(PC1) & 0.42 & 0.52 & 0.61 & 0.77 \\ 
 & UNB(PC2) & 0.66 & 0.59 & 0.68 & 0.79 \\ 
 & UNB(PC3) & 0.69 & 0.65 & 0.69 & 0.76 \\ 
 & UNB(PC4) & 0.48 & 0.51 & 0.69 & 0.79 \\ 
 & UNB(PC5) & 0.55 & 0.61 & 0.68 & 0.78 \\ 
 \cmidrule{2-6}
 & CTU & 0.77 & 0.77 & 0.86 & 0.90 \\ 
 \cmidrule{2-6}
 & MAWI & 0.86 & 0.68 & 0.42 & 0.62 \\ 
 \cmidrule{2-6}
 & TV\&RT & 0.95 & 0.98 & 0.90 & 0.74 \\ 
 \cmidrule{2-6}
 & GHom & 0.86 & 0.78 & 0.42 & 0.96 \\ 
 & SCam & 0.63 & 0.63 & 0.50 & 0.63 \\ 
 & SFrig & 0.96 & 0.95 & 0.57 & 0.94 \\ 
 & BSTch & 0.94 & 0.98 & 0.94 & 0.96 \\ 
\midrule

% \bottomrule
% \end{longtable}
% \end{scriptsize}
\end{tabular}
 \begin{tablenotes}
      \small
      \item *continue
    \end{tablenotes}
\end{threeparttable}
\end{table*}

\newpage
\begin{table*}[htbp!]
\centering
% \caption{AUCs for 4 ML approaches on basic feature representations with default parameters.}
% \label{tab:OCSVM_IF_AE_and_KDE_with_default_parameters} 
\begin{tabular}{|c|c|g|c|g|c|}
% \endfirsthead
% \endhead
% \toprule
\midrule
 Detector & Dataset & \Cell{STATS} & \Cell{SIZE} & \Cell{IAT}  & \Cell{SAMP-\\NUM}  \\ \midrule
\multirow{12}{*}{~\rule{0pt}{2.7ex}AE} & UNB(PC1) & 0.45 & 0.57 & 0.65 & 0.78 \\ 
 & UNB(PC2) & 0.64 & 0.58 & 0.68 & 0.81 \\ 
 & UNB(PC3) & 0.54 & 0.42 & 0.34 & 0.80 \\ 
 & UNB(PC4) & 0.80 & 0.41 & 0.37 & 0.77 \\ 
 & UNB(PC5) & 0.70 & 0.43 & 0.68 & 0.80 \\ 
 \cmidrule{2-6}
 & CTU & 0.29 & 0.72 & 0.86 & 0.83 \\ 
 \cmidrule{2-6}
 & MAWI & 0.39 & 0.48 & 0.43 & 0.60 \\ 
 \cmidrule{2-6}
 & TV\&RT & 1.00 & 1.00 & 0.95 & 0.76 \\ 
 \cmidrule{2-6}
 & GHom & 0.88 & 0.64 & 0.28 & 0.96 \\ 
 & SCam & 0.48 & 0.46 & 0.49 & 0.60 \\ 
 & SFrig & 0.94 & 0.72 & 0.72 & 0.96 \\ 
 & BSTch & 0.95 & 0.87 & 0.97 & 0.96 \\ 

\midrule
\multirow{12}{*}{~\rule{0pt}{2.7ex}KDE} & UNB(PC1) & 0.21 & 0.54 & 0.69 & 0.73 \\ 
 & UNB(PC2) & 0.56 & 0.60 & 0.79 & 0.78 \\ 
 & UNB(PC3) & 0.62 & 0.77 & 0.82 & 0.78 \\ 
 & UNB(PC4) & 0.37 & 0.62 & 0.86 & 0.74 \\ 
 & UNB(PC5) & 0.34 & 0.58 & 0.73 & 0.78 \\ 
 \cmidrule{2-6}
 & CTU & 0.75 & 0.73 & 0.76 & 0.92 \\ 
 \cmidrule{2-6}
 & MAWI & 0.66 & 0.49 & 0.39 & 0.58 \\ 
 \cmidrule{2-6}
 & TV\&RT & 1.00 & 1.00 & 0.95 & 0.67 \\ 
 \cmidrule{2-6}
 & GHom & 0.57 & 0.94 & 0.05 & 0.95 \\ 
 & SCam & 0.68 & 0.64 & 0.47 & 0.56 \\ 
 & SFrig & 0.94 & 0.91 & 0.74 & 0.95 \\ 
 & BSTch & 0.96 & 0.96 & 0.94 & 0.91 \\ 
\bottomrule
\end{tabular}
\end{table*}

\newpage
\begin{table*}[htbp!]
\centering 
\caption{AUCs for 2 additional ML approaches (GMM and PCA) on basic feature representations with default parameters.}
\label{tab:GMM_and_PCA_basic_representation_with_default_parameters} 
\begin{tabular}{|c|c|g|c|g|c|}
% \centering 
% \endfirsthead
% \endhead
\toprule
 Detector & Dataset & \Cell{STATS} & \Cell{SIZE} & \Cell{IAT} & \Cell{SAMP-\\NUM} \\ 
\midrule
\multirow{12}{*}{~\rule{0pt}{2.7ex}GMM} & UNB(PC1) & 0.36 & 0.22 & 0.62 & 0.79 \\ 
 & UNB(PC2) & 0.59 & 0.55 & 0.70 & 0.79 \\ 
 & UNB(PC3) & 0.63 & 0.11 & 0.74 & 0.80 \\ 
 & UNB(PC4) & 0.54 & 0.59 & 0.65 & 0.82 \\ 
 & UNB(PC5) & 0.81 & 0.41 & 0.70 & 0.78 \\ 
 \cmidrule{2-6}
 & CTU & 0.69 & 0.70 & 0.52 & 0.89 \\ 
 \cmidrule{2-6}
 & MAWI & 0.66 & 0.40 & 0.48 & 0.62 \\ 
 \cmidrule{2-6}
 & TV\&RT & 0.99 & 0.93 & 0.90 & 0.90 \\ 
 \cmidrule{2-6}
 & GHom & 0.39 & 0.51 & 0.67 & 0.96 \\ 
 & SCam & 0.63 & 0.62 & 0.42 & 0.63 \\ 
 & SFrig & 0.79 & 0.91 & 0.69 & 0.96 \\ 
 & BSTch & 0.94 & 0.08 & 0.93 & 0.96 \\ 
\midrule
\multirow{12}{*}{~\rule{0pt}{2.7ex}PCA} & UNB(PC1) & 0.33 & 0.21 & 0.30 & 0.72 \\ 
 & UNB(PC2) & 0.58 & 0.55 & 0.41 & 0.80 \\ 
 & UNB(PC3) & 0.62 & 0.15 & 0.29 & 0.79 \\ 
 & UNB(PC4) & 0.42 & 0.27 & 0.54 & 0.70 \\ 
 & UNB(PC5) & 0.39 & 0.22 & 0.39 & 0.75 \\ 
 \cmidrule{2-6}
 & CTU & 0.69 & 0.70 & 0.74 & 0.89 \\ 
 \cmidrule{2-6}
 & MAWI & 0.66 & 0.40 & 0.40 & 0.58 \\ 
 \cmidrule{2-6}
 & TV\&RT & 0.99 & 0.99 & 0.96 & 0.61 \\ 
 \cmidrule{2-6}
 & GHom & 0.40 & 0.42 & 0.06 & 0.96 \\ 
 & SCam & 0.63 & 0.62 & 0.60 & 0.57 \\ 
 & SFrig & 0.79 & 0.90 & 0.69 & 0.95 \\ 
 & BSTch & 0.97 & 0.08 & 0.92 & 0.95 \\ 
\bottomrule
\end{tabular}
\end{table*}

% \clearpage
 \newpage
\subsection{Effect of Fourier Domain Representation}

The figures below show differences in model accuracy for Fourier domain
representation vs. raw timeseries representations for different models and
datasets. Fourier domain representations in general do not improve model
accuracy.

\begin{figure}[htbp!]
\centering
\includegraphics[width=0.9\textwidth, height=18.cm]{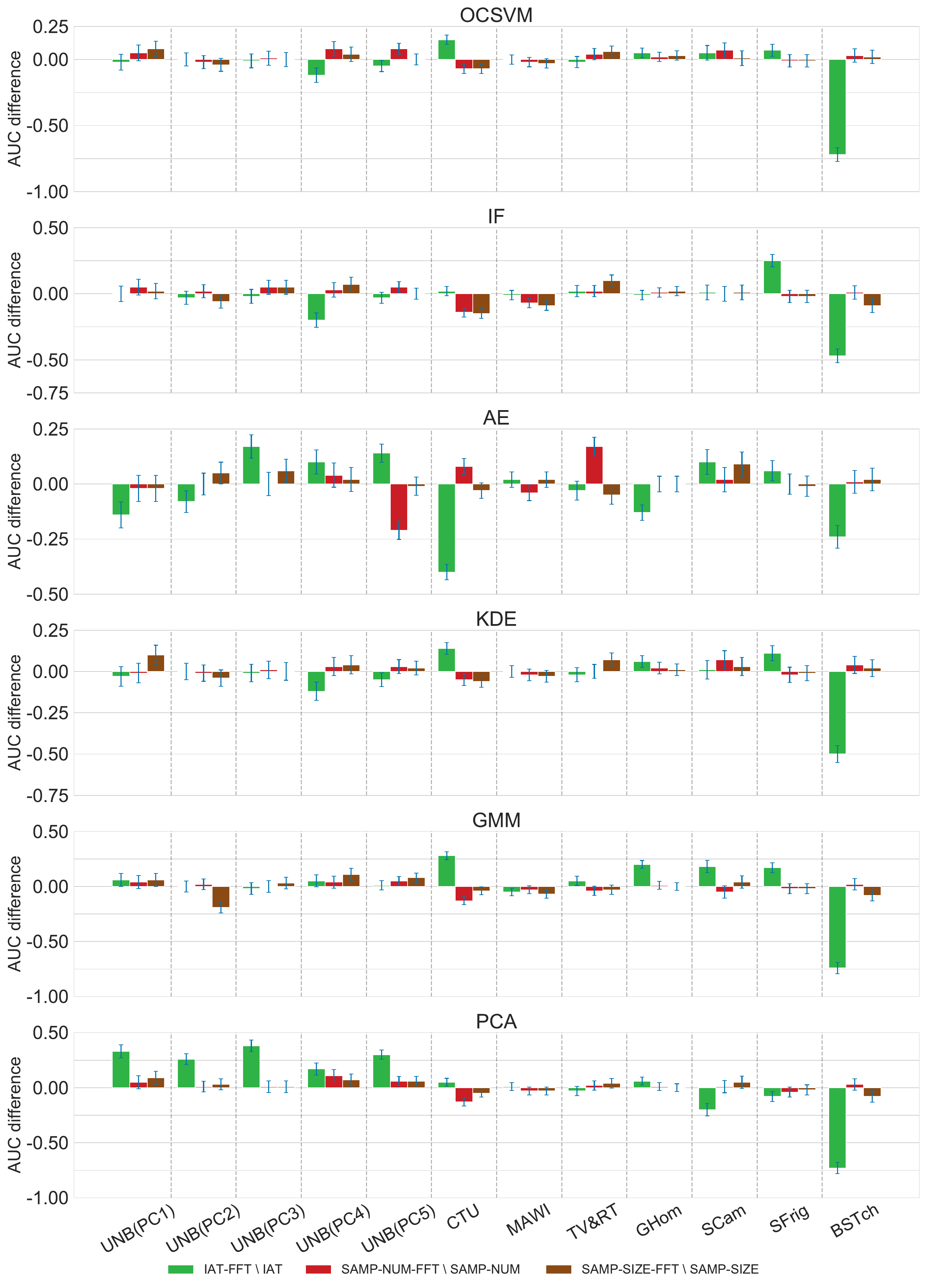}
\caption{Difference in AUC for {\bf FFT vs. raw time series} representations with default parameters.}
\label{fig:OCSVM_IF_AE_KDE_GMM_PCA-12-default-basic_representation}
\end{figure}

% \begin{figure}[htbp!]
% \centering
% \includegraphics[width=0.95\textwidth,height=8.63cm]{GMM_PCA-12-default-basic_representation.pdf}
% \caption{Difference in AUC for {\bf FFT vs. raw time series} representations generated by GMM and PCA with default parameters.}
% \label{fig:GMM_PCA-12-default-basic_representation}
% \end{figure}

\clearpage
\newpage
\subsection{Effect of Packet Size Information}

The figures below show differences in model accuracy 
as a result of including packet size information for different models and
datasets.  Packet size information generally improves model accuracy.

\begin{figure}[htbp!]
\centering
\includegraphics[width=0.9\textwidth,height=18.cm ]{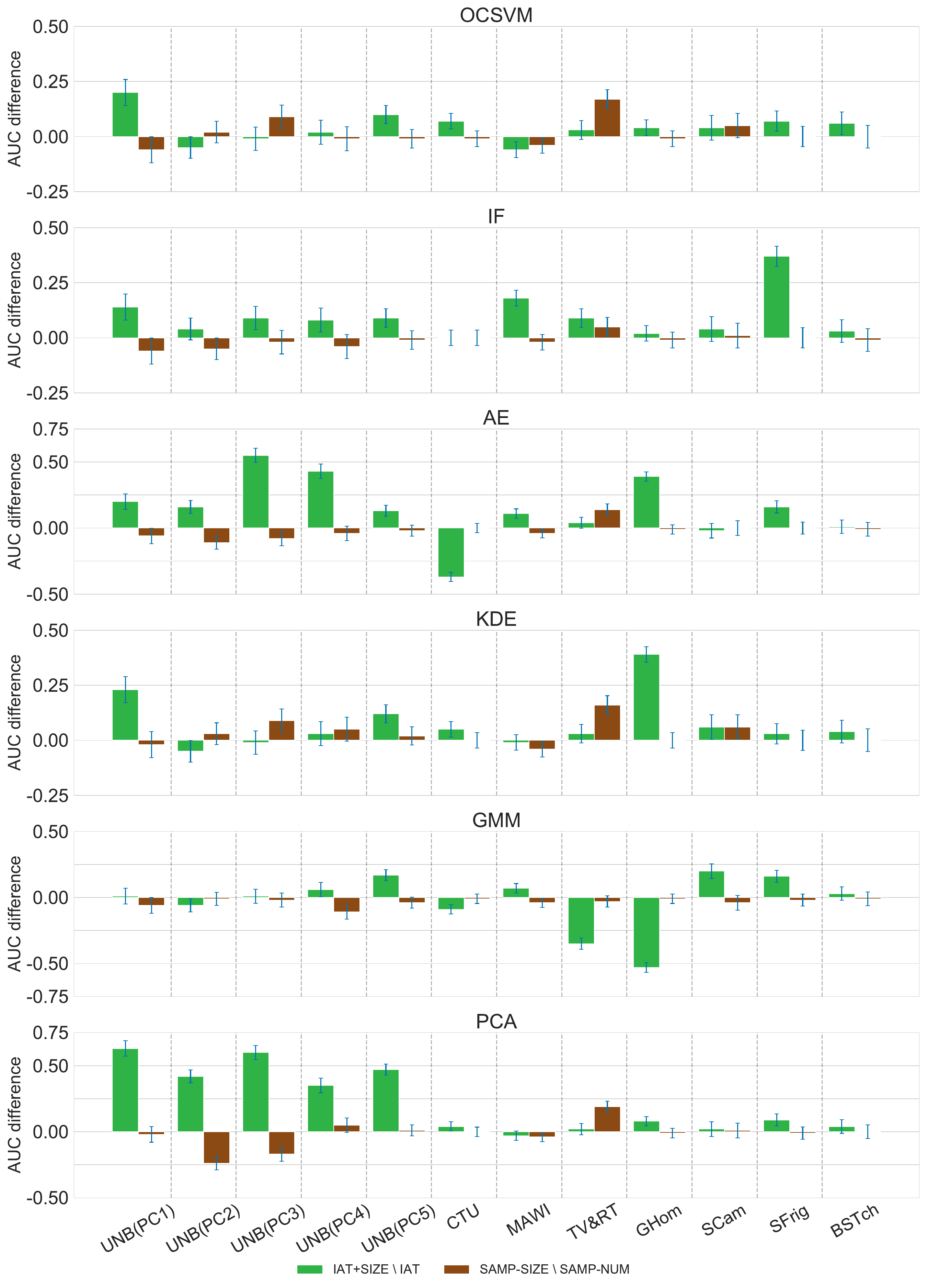}
\caption{Differences in AUC due to {\bf including vs. excluding packet size} information with default parameters.}
\label{fig:OCSVM_IF_AE_KDE_GMM_PCA-12-default-effect_size}
\end{figure}

% \begin{figure}[htbp!]
% \centering
% \includegraphics[width=0.95\textwidth,height=8.63cm]{GMM_PCA-12-default-effect_size.pdf}
% \caption{Differences in AUC due to {\bf including vs. excluding packet size} information generated by GMM and PCA with default parameters.}
% \label{fig:GMM_PCA-12-default-effect_size}
% \end{figure}

\clearpage
\newpage
\thispagestyle{plain} % removes running headers
\subsection{Effect of Packet Header}

The figures below show differences in model accuracy 
as a result of including packet header information for different models and
datasets.  Packet header information generally improves model accuracy.

\begin{figure}[htbp!]
\centering
\includegraphics[width=0.9\textwidth,height=18.cm ]{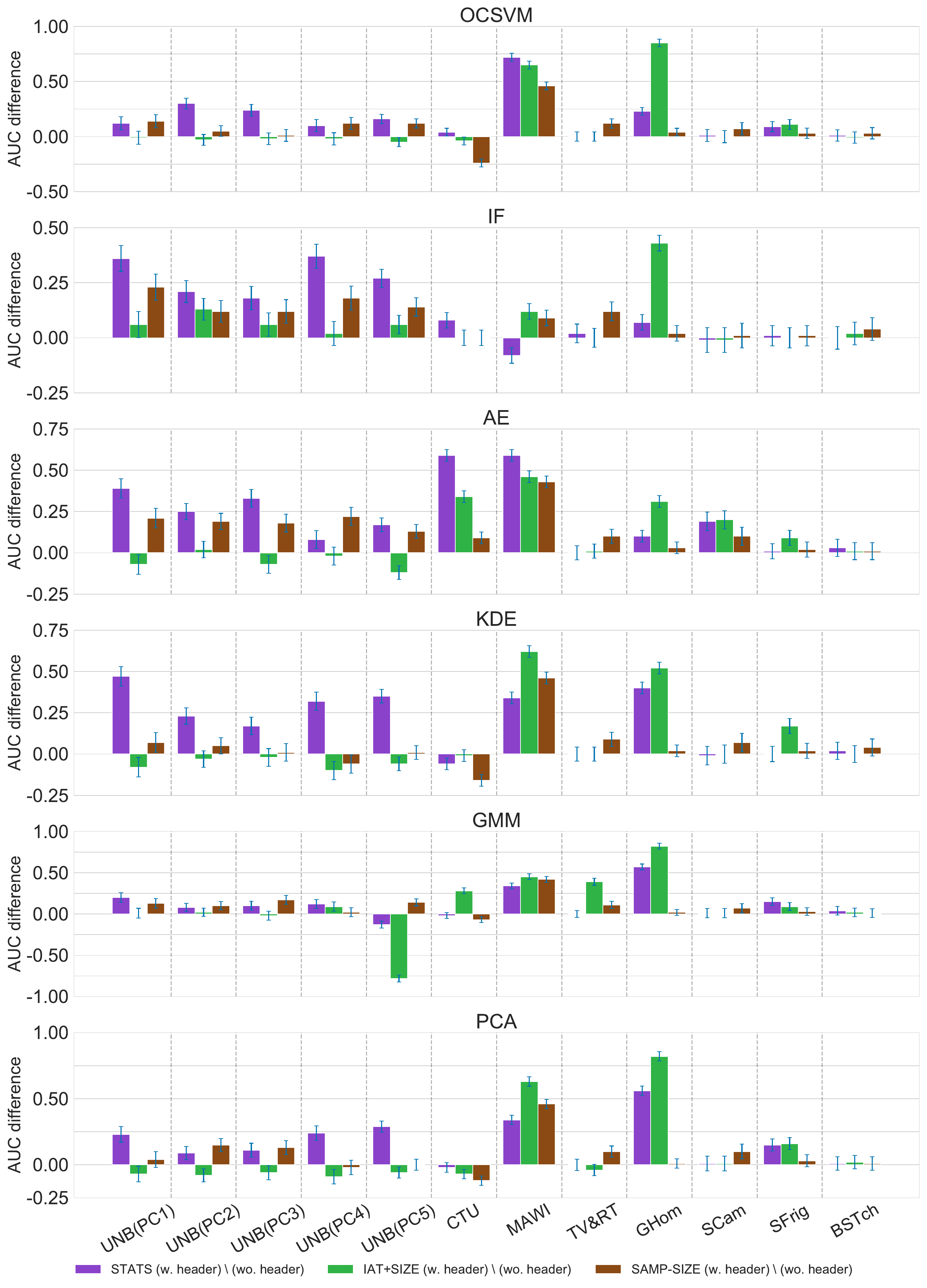}
\caption{Differences in AUC due to {\bf including vs. excluding packet header} information with default parameters.}
\label{fig:OCSVM_IF_AE_KDE_GMM_PCA-12-default-effect_header}
\end{figure}

% \begin{figure}[htbp!]
% \centering
% \includegraphics[width=0.95\textwidth,height=8.63cm]{GMM_PCA-12-default-effect_header.pdf}
% \caption{Differences in AUC due to {\bf including vs. excluding packet header} information generated by GMM and PCA with default parameters.}
% \label{fig:GMM_PCA-12-default-effect_header}
% \end{figure}